\documentclass[11pt,preprint,aaspp4]{aastex}
\accepted{September 25, 2001}
\journalid{565}{Astrophysics Journal}

\def\lsim{\lower 2pt \hbox{$\, \buildrel {\scriptstyle <}\over
         {\scriptstyle \sim}\,$}}

\begin{document} 
\title{GALACTIC POPULATIONS OF RADIO AND GAMMA-RAY PULSARS IN THE POLAR CAP 
MODEL}

\author{Peter L. Gonthier, Michelle S. Ouellette\altaffilmark{1} and Joel Berrier}
\affil{Hope College, Department of Physics, 27 Graves Place, Holland, MI
49422-9000}
\email{gonthier@physics.hope.edu, ouellette@pa.msu.edu, and bj277102@hope.edu}

\author{Shawn O'Brien}
\affil{University of Notre Dame, Department of Physics, Notre Dame, IN 46556}
\email{sobrien4@nd.edu}

\and
\author{Alice K. Harding}
\affil{NASA - Goddard Space Flight Center, Laboratory for High Energy
Astrophysics \\
Greenbelt, MD 20771}
\email{harding@twinkie.gsfc.nasa.gov}

\altaffiltext{1}{Michigan State University, National Superconducting
Cyclotron Laboratory, East Lansing, MI 48824-1116}

\begin{abstract} 
We simulate the characteristics of the Galactic population of radio and
$\gamma$-ray pulsars using Monte Carlo techniques.  At birth, neutron stars
are spatially distributed in the Galactic disk, with supernova-kick
velocities, and randomly dispersed in age back to $10^9$ years. They are
evolved in the Galactic gravitational potential to the present time. 
From a radio luminosity model, the radio flux is filtered through a
selected set of radio-survey parameters. $\gamma$-ray luminosities are
assigned using the features of recent polar cap acceleration models
invoking space-charge-limited flow, and a pulsar death valley further
attenuates the population of radio-loud pulsars.  Assuming a simple
emission geometry with aligned radio and $\gamma$-ray beams of 1 steradian
solid angle,
our model predicts that EGRET should have seen 7 radio-loud and 1
radio-quiet, $\gamma$-ray pulsars.  With much improved sensitivity, GLAST,
on the other hand, is expected to observe 76 radio-loud and 74
radio-quiet, $\gamma$-ray pulsars of which 7 would be identified as pulsed
sources.   We also explore the effect of
magnetic field decay on the characteristics of the radio and $\gamma$-ray pulsar
populations.  Including magnetic field decay on a timescale of 5 Myr 
improves agreement with the radio pulsar population and increases the predicted
number of GLAST detected pulsars to 90 radio-loud and 101 radio-quiet (9 pulsed)
$\gamma$-ray pulsars.  The lower flux threshold allows GLAST to detect 
$\gamma$-ray pulsars at larger distances than those observed by the radio surveys
used in this study.
\end{abstract} 



\section{Introduction}

With the advent of the {\it Compton Gamma Ray Observatory} (CGRO), the
number of $\gamma$-ray pulsars has grown to eight, with several additional
candidates.  We anticipate that many more pulsed sources will be added
to the list with the future telescope, {\it Gamma-Ray Large Area Space
Telescope} (GLAST) scheduled for launch in 2006.  The new body of
data with a much larger set of statistics will be essential in further
constraining pulsar models.  Many questions, such as the mechanism for
radio emission and its relationship to $\gamma$-ray emission, and the
location and geometry of the $\gamma$-ray emission currently remain
unanswered.  Among the known $\gamma$-ray pulsars, only Geminga is radio-quiet or
at least radio weak (Kuzmin \& Losovsky 1997 and Malofeev \& Malov 1997).
Geminga does not emit conventional radio emission that is detectable 
through the current radio surveys.  As a result, we refer to this object as
a radio-quiet, $\gamma$-ray pulsar.
Of the 271 sources listed in the Third EGRET
Catalog (Hartman et al. 1999), about 170 of these $\gamma$-ray point
sources have not been identified with sources at other wavelengths.  
Recently Grenier \& Perrot (1999) suggested that some of these unidentified
sources are correlated with the Gould Belt of massive stars from a
nearby Galactic structure consisting of an expanding disk of gas with
young stars ($\le$ 30 million years) inclined about $20^o$ to the
Galactic plane. Gehrels et al. (2000) found that the flux distribution
of 120 steady sources suggests two distinct groups: one having higher flux,
hard spectra and distributed along the Galactic plane, and the other
having lower flux, softer spectra and correlated with the Gould Belt.
The soft spectra and luminosity of this second group of sources are 
significantly different than those of the known $\gamma$-ray pulsars observed 
by EGRET.  Harding \& Zhang (2001) suggest that some of the sources 
associated with the Gould Belt are indeed radio-quiet, off-beam $\gamma$-ray 
pulsars seen at large angles to the magnetic pole. 

There are two main types of models proposed to explain pulsar high-energy
emission.  Polar cap models (Daugherty \& Harding 1996, Sturner et al. 1995) 
assume that particles accelerated above the neutron star polar caps 
produce $\gamma$-rays via curvature radiation or inverse Compton
scattering induced pair cascades in a strong magnetic field.  Outer
gap models (Cheng, Ho \& Ruderman 1986, Romani 1996, Hirotani \& Shibata
1999) assume that acceleration occurs along null charge surfaces in the
outer magnetosphere and that $\gamma$-rays result from photon-photon
pair cascades.  Polar cap and outer gap models predict different ratios of 
radio-loud to radio-quiet, $\gamma$-ray pulsars, primarily due to the different 
geometry of the high-energy emission region and its location relative to 
the radio emission region, thought to originate within tens of stellar
radii of the neutron star surface. 

Outer gap models generally predict small overlap of the radio and
high-energy pulsar populations, because the high-energy and radio pulses
that are visible to the same observer originate from opposite magnetic poles, 
and large numbers of radio-quiet,
$\gamma$-ray pulsars, because the outer gap beam is much larger than the
radio beam.  The study of Yadigaroglu \& Romani (1995), using the outer
gap model of Romani \& Yadigaroglu (1995), found that EGRET should
have detected about three times as many radio-quiet pulsars as
radio-loud pulsars in $\gamma$ rays.  Most of the pulsar population
should be seen only in $\gamma$ rays, as Geminga-like pulsars.
 The Monte-Carlo simulation of Cheng \& Zhang (1998) found that only
about $16\%$ of radio pulsars should be $\gamma$-ray pulsars.  They
expect 55 Geminga-like pulsars to have been detected by EGRET, about all of
the unidentified sources in the Galactic plane, as compared to 11 radio-loud
$\gamma$-ray pulsars expected.  The characteristics of the $\gamma$-ray
pulsar population, as seen by EGRET, in the inverse-Compton initiated polar cap
cascade model have been studied by Sturner \& Dermer (1996a).  They
predict that about $75\%$ of radio-selected pulsars should be
$\gamma$-ray pulsars but that only $25\%$ of $\gamma$-ray pulsars are
radio-quiet.  Although the uncertainties in these studies are large, it
is apparent that the differences between expected populations of polar
cap and outer gap models are larger than the uncertainties due to
model-dependent effects.

Since there has been no statistical study of the expected high-energy pulsar 
populations in the curvature radiation-induced polar cap cascade model 
(e.g. Daugherty \& Harding 1996), we have developed a Monte-Carlo code
similar to that of Sturner \& Dermer (1996a) to simulate the radio and 
$\gamma$-ray pulsar populations in the Galaxy. 
While we understand that the emission geometry is crucial to the
modeling of the $\gamma$-ray pulsar population, we present in this study a
simple geometric model that provides insight into the number of
radio-loud and radio-quiet, $\gamma$-ray pulsars detected by EGRET and those
expected to be detected by GLAST.  The simplest geometric assumption
suggested by the polar cap model is one in which the radio and the
$\gamma$-ray beams are aligned and equal in solid angle.  With this
assumption, we model the population of radio and $\gamma$-ray pulsars in
the Galaxy. We do not attempt to fit the observations.  Rather we use
standard distribution functions, evolution techniques, radio luminosity
models and recent $\gamma$-ray luminosity models suggested by an
acceleration-cascade model of the polar cap (Zhang \& Harding 2000). 
The simple model presented in this paper predicts a ratio of
radio-quiet to radio-loud, $\gamma$-ray pulsars (1/7) that is comparable to
the one observed by EGRET (1/8).  With a greater sensitivity, GLAST is
expected to observe a ratio of 74/76.  The large increase in the ratio
of radio-quiet to radio-loud, $\gamma$-ray pulsars is due to the greater
sensitivity of GLAST to detect $\gamma$-ray pulsars at greater distances
from the Earth than the radio surveys used in this study.  We plan to
have forthcoming the logical extension of this work that will invoke
more realistic emission geometries for both radio and $\gamma$-ray beams.

\section{Monte Carlo Simulation Model}

We develop a model to simulate the production of neutron stars within
the Galaxy, evolving their trajectories, periods and period derivatives
from their birth forward in time to the present.  We supply the radio
and $\gamma$-ray characteristics to each neutron star and filter its
properties through radio surveys and $\gamma$-ray thresholds (in and out of
plane) associated with EGRET and expected for GLAST.  These $\gamma$-ray
thresholds correspond to the number of photons required for the
instrument to identify the object as a point source. Higher thresholds
would be required to obtain sufficient photons to identify the object 
independently as a pulsed source.

\subsection{Initial Pulsar Period and Magnetic Field}

In most of our simulations, the magnetic field of the pulsar is assumed 
to be constant throughout its lifetime, although we also will explore the
effect of field decay.  A constant field requires that $(\dot P P)^{1/2}$ 
is equal to a constant
and, therefore, implies a nonzero $\ddot P$.  Following Bhattacharya et
al. (1992) and Gunn \& Ostriker (1970), we have assumed that the
magnetic field distribution can be represented by a Gaussian.  However,
we found it necessary to include two    additional Gaussians below the
main distribution to account more fully for the inferred distribution
from the period and period derivative of observed pulsars.  The pulsar's
surface magnetic field distribution is represented by the expression
\begin{equation}
\rho{ _B} =\sum\limits_{i=1}^3 {A_ie^{-\left( {\log B-\log B_i} \right)^2/\sigma _i^2}},
\label{eq:rhob}
\end{equation}
where the parameters take on the following values:
\begin{center}
\begin{tabular}{cccc}
&& Table 1 & \\ \hline
i & $A_i$ & $\log(B_i)$ & $\sigma_i$ \\ \hline
1 & 60. & 12.65 & 0.45 \\ 
2 & 2. & 11.9 & 0.6 \\
3 & .001 & 10.4 & 4 \\ \hline
\end{tabular}
\end{center}
An array normalized to unity is created with this distribution from
$\log(B) = 9.5$ to 13.5 in steps of 0.02, and used to randomly select
the pulsar's initial magnetic field.

The birth rate of neutron stars is assumed to be constant during the
history of the Galaxy.  Therefore, the age of the pulsar is randomly
selected from the present to $10^9$ years in the past.  While most of
the very old pulsars will not be observed due to their very long
periods, there will be some observed with ages close to $10^9$ years as
a result of their small magnetic fields and, therefore, small period
derivatives.  We use the expression by Shapiro \& Teukolsky
(1983) and Usov \& Melrose (1995) for a uniformly magnetized neutron
star where the inferred magnetic field is given by the star-centered
dipole relation, with magnetic moment $\mu = B_0R^3/2$,
\begin{equation}
B_{12}=6.4\ \times \ 10^7(P\dot P\ )^{1/2},
\label{eq:b12}
\end{equation}
where $B_{12}$ is in units of $10^{12}$ Gauss.  This relation is not
used in the pulsar catalog by Taylor, Manchester \& Lyne (1993) and in
Lyne \& Graham-Smith (1998) who instead use the approximate magnetic
dipole moment, $\mu = B_0R^3$.  Therefore, we multiplied the magnetic fields in the
pulsar catalog by a factor of 2 in order to make comparisons. 
Integrating this expression over time results in a pulsar period in
seconds at the present time by the expression:
\begin{equation}
P(t)=\left[ {P_o^2+1.54\times 10^{-8}B_{12}^2t} \right]^{1/2},
\label{eq:period}
\end{equation}
where $t$ is the age of the pulsar in years and the initial period $P_o$ is
in seconds. The initial period of the pulsar at birth is assumed to be
a fixed 30 ms.  The simulated distributions are rather insensitive to the
initial period as long as $P_o \lsim 100$ ms.  
Having the $P(t)$ along with the pulsar's magnetic field,
the pulsar's period derivative can be obtained assuming a braking index
of 3 from equation (\ref{eq:b12}).  Various studies (for example, Lyne, Pritchard \&
Smith 1988, Kaspi et al. 1994, and Boyd et al. 1995) measuring the
second period derivative suggest a lower braking index.  However,
systematics are difficult to obtain as second derivatives of pulsar
periods are very time consuming to measure.  At this time, we have not
attempted to model anything other than a braking index of 3.

Using the recent study of Zhang, Harding \& Muslimov (2000), we have
introduced a pulsar death line predicted by a multipole magnetic field
configuration near the stellar surface within a space-charge-limited
flow model and described by the expression
\begin{equation}
\log \dot P=2\log P-16.52.
\label{eq:pdot}
\end{equation}
The model includes the effect of general relativistic frame-dragging
discussed in Muslimov \& Tsygan (1992) and Muslimov \& Harding
(1997) essential in the development of the electric field parallel to
the magnetic field.  A more recent study of Harding \& Muslimov (2001)
has shown that the death line described by equation (\ref{eq:pdot}) 
roughly defines the boundary of pair production for a dipole field when 
non-resonant and resonant inverse Compton radiation processes are considered.   
Only pulsars with period derivatives greater than
those of equation (\ref{eq:pdot}) are further considered.

In addition to the death line, we have introduced a death valley in the
$\dot P - P$ space.  Otherwise, we find too many pulsars are predicted
near the death line that are clearly not present in the observed
distribution. In order to define the death valley, we have used a second
line from Zhang, Harding \& Muslimov (2000) that describes the death
line in the space-charged-limited flow model for a purely dipole
magnetic field near the stellar surface given by
\begin{equation}
\log \dot P=\left( {{5 \over 2}} \right)\log P-14.56.
\end{equation}
The above expression roughly describes the boundary of pair production 
for curvature radiation photons
in a pure dipole field (Harding \& Muslimov 2001). 
We have chosen an exponential decline in the pulsar population along
constant magnetic field in the $\dot P - P$ plane.  We
define a distance in $\dot P P$ from the multipole death line to the
location of the simulated pulsar in the death valley.  We randomly
select whether the particular pulsar is counted or rejected from further
consideration in the code.  The role of the death valley will be further
discussed in a later section. 

In addition to this case involving no magnetic field decay and a pulsar
death valley, we have simulated a case in which the field is allowed to decay
exponentially with time constant, $\tau_D$, given by the expression
\begin{equation}
B(t)=B_{o12}e^{-t/\tau_D},
\label{eq:bdec}
\end{equation}
where $\rm B_{o12}$ is the magnetic field of the pulsar at birth.  In the field 
decay case, we found that the pulsar death valley is no longer required.
Assuming magnetic dipole spin-down and initial period, $\rm P_o$, the period
and the period derivative at the present time can be obtained by
\begin{eqnarray}
P^2&=&P^2_o+7.69\times10^{-9}B^2_{o12}\tau_D(1-e^{-t/\tau_D}),\ {\rm and} \nonumber \\
\dot P&=&2.44\times 10^{-16}e^{2t/\tau_D}{B^2_{o12} \over P},
\label{eq:pdec}
\end{eqnarray}
where P and $\rm P_o$ are in seconds and t and $\tau_D$ are in years.
Due to the field decay, the spin-down age of the observed pulsars has to be 
determined using
the equation
\begin{equation}
Age={\tau_D \over 2}\ln\left[{3.17\times10^{-8}P\over {\dot P\tau_D}}+1\right].
\label{eq:age}
\end{equation}
In the limit as $\tau_D$ goes to infinity, the age becomes the traditional 
characteristic age of a pulsar given by $P/{2\dot P}$. 
Due to the field decay, we find that a single Gaussian is sufficient to
describe the initial magnetic field distribution at birth of the majority 
of the pulsar population in the $\dot P - P$ space.  The values are
listed in Table 2 for the Gaussian parameters in equation (\ref{eq:rhob}).

\begin{center}
\begin{tabular}{cccc}
&& Table 2 & \\ \hline
i & $A_i$ & $\log(B_i)$ & $\sigma_i$ \\ \hline
1 & 1.0 & 12.75 & 0.4 \\ \hline
\end{tabular}
\end{center}
The role of field
decay will be discussed in a later section along with the results.

\subsection{Spatial Distribution of Pulsars}
In a cylindrical coordinate system with the origin at the Galactic
center, we assume that the birth location of neutron stars is well
described by the following distributions as indicated in Paczy\'{n}ski
(1990):
\begin{eqnarray}
\rho _z(z)dz&=&e^{-|z|/z_{\exp }}/z_{\exp }dz,\ {\rm and} \nonumber \\
 \rho _R(R)dR&=&a_Re^{-R/R_{\exp }}R/R_{\exp }^2dR,
\end{eqnarray}
where $z$ is the distance from the Galactic disk and $R$ is the distance
from an axis through the Galactic center perpendicular to the Galactic
disk and is given by:
\begin{equation}
R^2=x^2+y^2.
\end{equation}
The constants are defined as
\begin{eqnarray}
a_R&=&[1-e^{-R_{\max }/R_{\exp }}(1+R_{\max }/R_{\exp })]^{-1},\nonumber \\
  R_{\exp }&=&4.5\ {\rm kpc},\nonumber \\
  z_{\exp }&=&0.075\ {\rm kpc},\ {\rm and} \\
  R_{\max }&=&20\ {\rm kpc}. \nonumber
\end{eqnarray}
Defined in this fashion the following integrals are normalized to unity,
\begin{eqnarray}
\int\limits_0^\infty  { e^{-|z|/z_{\exp } }/z_{\exp }dz } = 1,\ {\rm and}\nonumber \\
 \int\limits_0^{R_{\max }} { a_Re^{-R/R_{\exp }}R/R_{\exp }dR } = 1.
\end{eqnarray}
With these distributions, $\rho_z$, and $\rho_R$, the initial position
of the neutron stars can be chosen using a random number, $\cal R$
\begin{equation}
z=-\ln (1-{\cal R})z_{\exp },
\end{equation}
with the sign of z chosen by a second random number.  However, as
the inversion of the integral of the $\rho_R$ function involves solving a transcendental 
equation, the distance $R$ is chosen randomly by creating a normalized
array,
\begin{equation}
I_R(R)=e^{-R/R_{\exp }}R,
\end{equation}
and performing a linear interpolation between neighboring values.  The
azimuthal angle, $\phi$, is chosen randomly between 0 and 2$\pi$.
\subsection{Supernova Kick Velocity Distributions} The initial velocity
distribution given to a neutron star during a supernova has been the
subject of much discussion.  The three-dimensional space velocities of
neutron stars described by Lyne \& Lorimer (1994) have been obtained
from a two-dimensional distribution, $x^{0.3}\over {1+x^{3.3}}$ (see
also Mollerach \& Roulet (1997)) where $x$ is proportional to the
transverse velocity.  The three-dimensional space velocity distribution
can be accurately represented by an expression from Sturner \& Dermer
(1996b) and has the form
\begin{equation}
\rho _v(\zeta )={4 \over \pi }\left( {{\zeta  \over {1+\zeta ^4}}} \right),
\end{equation}
where $\zeta=v/350\ {\rm km/s}$ and $v$ is the random three-dimensional
velocity of the neutron star in its rest frame distributed
isotropically.  The most probable and mean $\zeta$ from this
distribution are 0.76 and 1.41, respectively.  The advantage of this
function is that it is not only normalized, but the integral can be
easily inverted to obtain the velocity directly from a random number
with the expression
\begin{equation}
\zeta =\sqrt {\tan \left( {{{\pi{\cal R} } \over 2}} \right)}.
\end{equation}
As discussed later in the text, we have chosen this functional form, but
we have used instead $\zeta=v/120\ {\rm km/s}$ in order to obtain better
agreement with the $z$ distribution of the pulsars.

\subsection{Galactic Gravitational Potentials}
We adopt the negative of the potential functions as defined in
Paczy\'{n}ski (1990),
\begin{eqnarray} 
&\Phi _i(R,z)= \mbox{\Large $
{{-GM_i} \over {\left\{ {R^2+\left[ {a_i+\left( {z^2+b_i^2} 
\right)^{1/2}} \right]^2} \right\}^{1/2}}} $},\ {\rm and}\nonumber \\
  &\Phi _h(r)=\mbox{\Large $
{-{GM_c} \over {r_c}}\left[ {1+{1 \over 2}\ln \left( {1+{{R_h^2} \over {r_c^2}}} \right)} {-{1 \over 2}\ln \left( {1+{{r^2} 
  \over {r_c^2}}} \right)-{{r_c} \over r}\tan ^{-1}\left( {{r \over {r_c}}} 
  \right)} \right] $}, 
\end{eqnarray}
where $\Phi_1$ corresponds to the spheroid, $\Phi_2$ corresponds to the
disk potentials and $\Phi_h$ is the halo potential.  We use the same
choice of parameters as in Paczy\'{n}ski (1990)
\begin{eqnarray}
a_1=0, & b_1 =0.277\ {\rm kpc}, & M_1 =1.12\times 10^{10}M_{\odot}, \nonumber \\
  a_2=3.7\ {\rm kpc}, & b_2=0.20\ {\rm kpc},&M_2=8.07\times 10^{10}M_{\odot}, \\
  & r_c=6.0\ {\rm kpc}, & M_c=5.0\times 10^{10}M_{\odot}. \nonumber
\end{eqnarray}
where the variable, $r$, corresponds to the radial distance from the Galactic
center:
\begin{equation}
r=\sqrt {x^2+y^2+z^2.}
\end{equation}
We have added a constant term to the halo potential that has been left out of the expression
in Paczy\'{n}ski (1990).  The constant negative term dominates over
the $r$ dependent term resulting in an overall negative potential.  In the constant term, $R_h$ is
the radius of the halo with a typical value of 41 kpc (Binney \& Tremaine 1987). However, this term does not affect
the pulsar trajectories as the equations of motion depend on the derivatives of the potential.  It is
important, if one is interested in the total energy to determine if the stars are unbound.  Of course, 
if the star escapes the halo, the potential would change.  The pulsars of interest 
in this study are well within the halo, and, therefore, this form of the potential is appropriate to
evolve the trajectories.  
The Lagrangian in units of energy per unit mass has the form
\begin{equation}
L=T-U={1 \over 2}\left( {\dot R^2+R^2\dot \phi ^2+\dot z^2} \right)-\Phi (R,z),
\end{equation}
where $\Phi(R,z)$ is the total potential energy per unit mass. 
 Since the Lagrangian is independent of $\phi$, the angular momentum,
$\ell$, along the $z$ direction is a constant of motion.
\begin{eqnarray}
& \mbox{\Large $
{{\partial L} \over {\partial \dot \phi }}=R^2\dot \phi =\ell $} ,{\rm \ so\ that} \nonumber \\
  & \mbox{\Large $
v_{cir}=R\dot \phi ={\ell  \over R} $}.
\end{eqnarray}
The circular velocity, $v_{cir}$, is the motion of the star in the
Galactic plane (for $z = 0$).  The equations of motion to be integrated
for the $R$ and $z$ directions have the forms
\begin{eqnarray}
&\ddot R=\mbox{\Large $
{{\ell ^2} \over {R^3}}-{{\partial \Phi } \over {\partial R}} $},
\ {\rm and} \nonumber \\
  &\mbox{\Large $
\ddot z=-{{\partial \Phi } \over {\partial z}} $}.
\end{eqnarray}
Of course the supernova explosion will impart an initial random velocity 
(eqn [16])
that needs to be taken into account in determining the initial angular
momentum and the total energy of the system in its Galactic orbit.
Once the initial conditions are established, the
trajectories are integrated using a fourth order Runge-Kutta routine
designed to maintain a high level of accuracy in the conservation of
total energy of one part per $10^8$ during the trajectory from its birth to the present time.

\subsection{Radio Luminosity}
We obtain the radio luminosity of the pulsars at 400 MHz, $L_{400}$,
using the radio pulsar model of Narayan \& Ostriker (1990) where the
normalized luminosity distribution is dithered by a function given by
\begin{equation}
\rho _{L_{400}}=0.5\lambda ^2e^{-\lambda },
\label{eq:dither}
\end{equation}
where
\begin{equation}
\lambda =3.6\left( {\log \left( {{{L_{400}} \over {\left\langle {L_{400}} 
\right\rangle }}} \right)+1.8} \right),
\label{eq:lamb}
\end{equation}
and where the average radio luminosity, $\langle L_{400}\rangle$, is
given by
\begin{equation}
\log \left\langle {L_{400}} \right\rangle =6.64+{1 \over 3}\log {{\dot P} \over {P^3}}.
\label{eq:l400}
\end{equation}
This is the luminosity law that was adopted by Bhattacharya et al. (1992) along with the 
above dithering function.  This luminosity function derived from $P$ and $\dot P$ was 
obtained by Pr\'{o}szy\'{n}ski \& Przybicie\'{n} (1984). 
We build an array with the distribution indicated by equation (\ref{eq:dither}) from
$\lambda$ = 0 to 20 in steps of 0.02, and using a random number, we
linearly interpolate between neighboring array elements to select a
$\lambda$.  Given the period and period derivative, equation (\ref{eq:l400}) gives
the average luminosity at 400 MHz, $\langle L_{400}\rangle$, and
together with equation (\ref{eq:lamb}) provides the luminosity of the pulsar at 400
MHz, $L_{400}({\rm mJy\cdot kpc^2})$. However, as discussed later, we find
that we had to modify the intercept of equation (\ref{eq:l400}) in order to achieve better
agreement with the observed pulsar distributions.  The flux in mJy is then
obtained from
\begin{equation}
S_{400}={{L_{400}} \over {\Delta \Omega d^2 }},
\label{eq:s400}
\end{equation}
where $d$ is the distance from the Earth in kpc.  We have assumed a
constant solid angle of $\Delta\Omega=1$ steradian.  The simulated flux
at 400 MHz is scaled to the observing frequency of the surveys we model
using a spectral index of $-1.7$. This is the average spectral index in
the frequency range between the fluxes of $S_{400}$ and $S_{1400}$ of
the select group of observed pulsars and is in agreement with Johnston
et al. (1992). Pulsars with fluxes greater than the survey flux
threshold are detected.

\subsection{Radio Detection}
The sensitivity of a particular survey is a
function of several parameters usually given by the expression (Dewey et
al. 1985) for
$S_{\min}\ (mJy)$,
\begin{equation}
S_{\min }={{C_{thres}\left[ {T_{rec}+T_{sky}(l,b)} \right]} \over {G
\sqrt {N_pBt}}}\sqrt {{W \over {P-W}}},
\label{eq:smin}
\end{equation}
where $C_{thres}$ is the detection threshold S/N, $T_{rec} ({\rm K})$ is
the receiver noise temperature, $T_{sky}(\ell,b) ({\rm K})$ is the sky
temperature in the direction being searched, G (K/Jy) is the telescope
gain, which we adjust to reflect various system losses, $N_p$ is the
number of polarizations (usually 2), $B$ (MHz) is the total bandwidth,
$t$ (s) is the integration time, $P$ (ms) is the pulsar period and $W$
(ms) is the effective pulse width. 

Expressions for $S_{\min}$ in the literature sometimes (Johnson et al.
1992, Lyne et al. 1998, Manchester et al. 1996) include an additional
term, $\beta$, in the numerator to explicitly account for the losses due
to digitization and other system losses.  In such cases, we incorporate
the reported values of $\beta$ by adjusting the value of $G$ (Table 4),
increasing $S_{\min}$ by effectively reducing the gain. We use the sky temperature at 408
MHz determined by Haslam et al. (1982) that has been made available in
machine readable form (http://skyview.gsfc.nasa.gov).  Given the
Galactic longitude and latitude, we interpolate a $512\times 1024$ table
of sky temperatures at 408 MHz. We then scale the temperature to the
observing frequency $\nu$ (MHz) through a power law given by (Johnson et
al. 1992) in the form
\begin{equation}
T_{sky}(\nu )=T_{sky,408}\left( {{{408\ {\rm MHz}} \over \nu }} \right)^{2.6}.
\end{equation}
The effective pulse width $W$ is given by
\begin{equation}
W^2=W_o^2+\tau _{samp}^2+\tau _{DM}^2+\tau _{scat}^2+\tau_{trail DM},
\label{eq:pwid}
\end{equation}
where $W_o$ is the intrinsic pulse width (here assumed to be
$0.05P$(ms)), $\tau_{samp}$ represents low pass filter time-constant
applied before sampling or, if unknown, an assumed value at twice the
sampling interval ($2\Delta t$), $\tau_{DM}$ is the dispersion smearing
time over one frequency interval, $\Delta\nu$ (MHz), $\tau_{scat}$ (ms)
is the time broadening of the pulse due to interstellar scattering.  The
dispersion broadening time, $\tau_{DM}$ (ms), across one frequency
channel, $\Delta \nu$ is related to the dispersion measure (DM) and has
the form
\begin{equation}
\tau _{DM}=8.3\times 10^6{\Delta\nu DM \over {\nu^3}}.
\end{equation}
The dispersion measure $DM ({\rm pc/cm^3})$ is obtained using the Taylor
\& Cordes (1993) distance model, where we have translated the FORTRAN
routine into C code and used an extended trapezoidal integration routine
(Press et al. 1992).  The distance model also provides the scattering measure
SM ($\rm kpc\cdot m^{-20/3}$), which allows one to obtain the broadening
time due to interstellar scattering, $\tau_{scat}$ (ms), used in
equation (\ref{eq:pwid}) with the expression 
\begin{equation}
\tau_{scat} = 1000{\left(SM\over 292\right)}^{1.2} d {\left(\nu\over
1000\right)}^{-4.4}, 
\end{equation} 
where the scattering time is scaled (Johnston et al. 1992) from 1 GHz
assumed in the Taylor \& Cordes (1993) model to the observing frequency,
$\nu$.  The final term in (\ref{eq:pwid}), $\tau_{trail DM}$,
corresponding to the fourth term in the expression for $W$ in Dewey et al.
(1985), is the additional time broadening when the sampling is performed
at a DM different than the actual DM of the pulsar.  This term becomes
important for low period pulsars.  Since we are interested in pulsars
with periods greater than 30 ms, we have neglected this term in our
simulations.  As will be indicated later, we do not see many pulsars
with periods less than 100 ms where this term might add an important
contribution to the smearing time, an effect that we will consider
including in subsequent refinements of the model presented here.

\sloppy{We
have selected a set of eight surveys to filter our simulations in making comparisons
with the 707 pulsars of the Princeton catalog (Taylor, Manchester \& Lyne 1993)
from \underline{http://pulsar.princeton.edu/pulsar/catalog.shtml}.  We selected
pulsars from the catalog with positive period derivatives, periods greater than
30 ms and pulsars that are not in globular clusters and are not in binary
systems to form a comparison set of 445 pulsars observed by the indicated
surveys in Table 3.  This group includes 90\% of the 496 pulsars having these
characteristics and observed by all the surveys in the Princeton catalog.  We
have multiplied by a factor of 2 the magnetic field strengths of the pulsars in
the catalog to compare with our adopted surface field in equation
(\ref{eq:b12}). 
\begin{center}
\begin{tabular}{cllclcc}
\multicolumn{7}{c}{Table 3}\\ \hline
\multicolumn{7}{c}{Selected Pulsar Surveys} \\ \hline
Survey & Name & Reference & 				Eff. 		  	  & \multicolumn{3}{c}{Boundaries} \\ \hline 
       &      &           &                                   & Type & $\alpha/\ell$ & $\delta/ b$ \\ \hline
1	& Molonglo 2	& Manchester et al. (1978) 	  & 1.00          & Eq.  & [0,360]  & [-85,20]   \\
2	& Green Bank 2	& Dewey et al. (1985) 	      & 0.22          & Eq.  & [0,360]  & [-18,90]   \\
3	& Green Bank 3	& Stokes et al. (1986) 	  & 0.17          & Eq.  & [0,360]  & [-18,90]   \\ 
4	& Arecibo 2	& Stokes et al. (1986) 		  & 0.58          & Gal.  & [40,65]  & [-10,10]   \\ 
5	& Parkes 1	& Johnson et al. (1992) 	  & 0.91          & Gal.  & [270,20] & [-4,4]     \\
6	& Arecibo 3	& Nice et al. (1993) 		  & 0.49          & Gal.  & [35,65]  & [-8.0,8.0] \\ 
7	& Parkes 2	& Manchester et al. (1996) 	  & 0.95          & Eq.  & [0,360]  & [-90,0]    \\
    &           & Lyne et al. (1998) 		  &               &      &          &            \\ 
8   & Jodrell Bank 2 & Clifton \& Lyne (1986) & 0.93          & Gal.  & [-5,105] & [-1,1]     \\ 
    &                & Clifton et al. (1992)  &               &      &          &            \\ \hline
\end{tabular}
\end{center}
The essential parameters (Z. Arzoumanian, priv. comm.) of these surveys
are indicated in Table 4 and discussed in the text.  The gain for the
Jodrell Bank 2 (A. Lyne, priv. comm.) has been adjusted to reproduce the
long-period $S_{\min}$ in Clifton et al. (1992).  We also do not adjust
the gain as a function of zenith angle for the Arecibo surveys.
\begin{center}
\begin{tabular}{cccccccccc}
\multicolumn{10}{c}{Table 4} \\ \hline
\multicolumn{10}{c}{Survey Parameters} \\ \hline
\scriptsize Survey	&\scriptsize  Gain	& $C_{Thres}$ & $T_{rec}$ & 
$\nu$ & $t$ & $\Delta t$ & $\tau$ & $B$ & $\Delta \nu$ \\ 
& \scriptsize (K/Jy) & \scriptsize & \scriptsize (K) & \scriptsize (MHz) &
\scriptsize (s) & \scriptsize (ms)  & \scriptsize (ms) & \scriptsize (MHz) & 
\scriptsize (MHz) \\ \hline 
1	& 5.10	& 5.4	& 210	& 408	& 40.96	& 20	& 40   & 3.2	& 0.8 \\
2	& 0.89	& 7.5	& 30	& 390	& 137	& 16.7	& 33.5 & 16.	& 2.00  \\ 
3	& 0.95	& 8.0	& 30	& 390	& 131	& 2.0	& 2.2  & 8.0	& 0.25  \\
4	& 10.9	& 8.0	& 90	& 430	& 39.3	& 0.30	& 0.4  & 0.96	& 0.06	\\ 
5	& 0.24	& 8.0	& 45	& 1520	& 157	& 1.20	& 2.4  & 320	& 5.0	\\
6	& 13.35	& 8.5	& 75	& 430	& 68.2	& 0.52	& 0.5  & 10.0	& 0.078	\\ 
7	& 0.43	& 8.0	& 50	& 436	& 157	& 0.30	& 0.6  & 32.0	& 0.125	\\ 
8   & 0.40  & 6.0   & 40    & 1400  & 524   & 2.00  & 4.0  & 40     & 5     \\
\hline
\end{tabular}
\end{center}
These parameters are used to calculate a minimum radio flux from
equation (\ref{eq:smin}) that is compared to the calculated radio flux
from equation (\ref{eq:s400}) scaled to the observing frequency with a
spectral index of $-1.7$.  The efficiency of the survey (Eff. in Table
3) has been obtained from the number of survey beams times the reported
solid angle at half-power beamwidth, divided by the area enclosed by the
survey boundaries on the sky. The efficiency is assumed to be constant
over the area surveyed.  This is not the case for some of the surveys
where more sampling was done in certain regions while other regions were
sampled more sparely.  In Figure 1, we indicate the flux thresholds for
the different surveys used in the simulation assuming a sky temperature
of 200 K at 408 MHz characteristic of the Galactic disk and a dispersion
measure of 200 ${\rm cm^{-3}\cdot pc}$. We realize that for some surveys
that do not cover this Galactic disk, this temperature might not be
appropriate.  We choose a common temperature for the sake of comparing
the $S_{\min}$ of each of the surveys at the corresponding observing
frequency.  However, in the actual simulation the sky temperature at 408
MHz is obtained using the all-sky map of Haslam et al. (1982) and, then
scaled to the particular observing frequency of the survey being tested.
For the calculation of $S_{\min}$ in Figure 1 only, we assume a
$\tau_{scat}$ obtained from the formula by Bhattacharya et al. (1992)
scaled to the observing frequency and given by
\begin{equation}
\tau_{scat} = \left[10^{-4.62 + 1.14 \log(DM)} + 10^{-9.22 + 4.46 \log(DM)}\right]\left({400\over{\nu}}\right)^{4.4}.
\end{equation} 
In the actual simulation, we obtain the scattering measure, $SM$, from
the Taylor \& Cordes (1993) distance model as indicated above. In Figure
1, The smallest flux thresholds occur for large periods where the
intrinsic pulsar width, $W_o$, dominates over other terms in equation
(\ref{eq:pwid}) resulting in a constant $S_{\min}$ from equation
(\ref{eq:smin}). The flux threshold increases with decreasing period
when the intrinsic pulse width is dominated by the other pulse
broadening terms in equation (\ref{eq:pwid}) and also as a result of the
period dependence in equation (\ref{eq:smin}). The Green Bank 2 and
Molonglo 2 surveys are not very sensitive to pulsars with periods
smaller than 0.1 seconds.

\subsection{$\gamma$-ray Luminosity}
We have taken the expressions describing the $\gamma$-ray luminosity from
the work of Zhang \& Harding (2000) where a polar-cap model simulates
the pair cascade region, with curvature radiation of the primary particles 
and synchrotron radiation and inverse Compton scattering of
subsequent higher generation pairs.  In addition, the model uses the
self-consistent acceleration model of Harding \& Muslimov (1998) to
produce the primary particles.  The $\gamma$-ray luminosity is constrained
by equation (61) of Zhang \& Harding (2000) to be less than the
spin-down luminosity, or
\begin{equation}
B_{p,12}P^{-7/4}(\cos \alpha )^{-5/4}\ge 1.65,
\label{eq:regi}
\end{equation}
where we have used $\alpha=30^o$.  If equation (\ref{eq:regi}) is
not satisfied, the $\gamma$-ray luminosity is equal to the spin-down
luminosity,
\begin{equation}
L_{sd}=9.68\times 10^{30}B_{p,12}^2P^{-4}I_{45},
\end{equation}
where $I_{45}$ is the moment of inertia in $10^{45} {\rm g \cdot cm^2}$.  If
equation (\ref{eq:regi}) is satisfied, then the pulsar is in regime I if
\begin{equation}
B_{p,12}^{1/7}P^{-9/28}(\cos \alpha )^{3/28}>3.0,
\end{equation}
otherwise the pulsar is in regime II.  The $\gamma$-ray luminosities for
the two regions are then given by the forms
\begin{eqnarray}
L_\gamma (I)=4.8\times 10^{31}\ {\rm erg\cdot s^{-1}}\ B_{p,12}^{6/7}P^{-27/14}
(\cos\alpha )^{8/7},\ {\rm and} \nonumber \\
L_\gamma (II)=1.6\times 10^{31}\ {\rm erg\cdot s^{-1}}\ B_{p,12}^{}P^{-9/4}
(\cos \alpha )^{5/4}.
\label{eq:Lgamma}
\end{eqnarray}

These expressions correspond to the equations (59) and (60) in the work
by Zhang \& Harding (2000) with an improved form for the luminosity in
region I based on a more accurate expression for $E_{||}$.  Equation 
(\ref{eq:Lgamma}) reproduces the dependence of $\gamma$-ray luminosity 
on $\dot E^{1/2} \propto B_p P^{-2}$ observed for CGRO detected pulsars 
(Thompson et al. 1997), and also agrees with the empirical luminosity law
for CGRO pulsars found by McLaughlin \& Cordes (2000).  Having the
$\gamma$-ray luminosity, we can compute the flux assuming a solid angle
$\Delta\Omega$ by the form
\begin{equation}
S_\gamma ={{L_\gamma } \over {\Delta \Omega d^2}},
\end{equation}
where we have assumed the same solid angle of 1 steradian as in the case
for the radio emission.  {\bf We have used the flux thresholds (D. Thompson, priv. comm.) for
in and out of the Galactic plane as indicated in Table 5.}

\begin{tabular}{ccc}
\multicolumn{3}{c}{Table 5} \\ \hline
\multicolumn{3}{c}{In and Out-of-Plane Flux Thresholds} \\ \hline
	& EGRET	& GLAST \\
In-plane $\|b\|<10^o$	& $1.6\times 10^{-7} \ {\rm photons/(cm^2\cdot s)}$	& 
$5.0\times 10^{-9}\ {\rm photons/(cm^2\cdot s)}$ \\
Out-of-plane $\|b\|\ge 10^o$ & $7.0\times 10^{-8}\ {\rm photons/(cm^2\cdot s)}$	& 
$2.0\times 10^{-9}\ {\rm photons/(cm^2\cdot s)}$ \\ \hline
\end{tabular}

\subsection{$\gamma$-ray Spectra}
The number of simulated $\gamma$-ray pulsars depends crucially on the
average $\gamma$-ray energy assumed in the calculation that converts the
luminosity from erg/s to a flux $({\rm photons\cdot cm^{-2}\cdot s^{-1})}$ which gets
compared to the instrument flux thresholds.  If the $\gamma$-ray spectrum
is assumed to be a power law of the form
\begin{equation}
{{dN} \over {dE_\gamma }}=AE_\gamma ^{-\alpha },
\end{equation}
where $A$ is a normalization factor and $\alpha$ is the spectral index, the average energy
above an energy threshold, $\epsilon_{th}$, can be obtained by
\begin{equation}
\left\langle {E_\gamma } \right\rangle =\left( {{{\alpha -1} \over {\alpha -2}}} \right)
\left( {\varepsilon _{\max }^{}\varepsilon _{th}^{}} \right){{\varepsilon _{th}^
{\alpha -2}-\varepsilon _{\max }^{\alpha -2}} \over {\varepsilon _{th}^{\alpha -1}-
\varepsilon _{\max }^{\alpha -1}}},
\label{eq:egam}
\end{equation}
where $\epsilon_{max}$ is the high energy cutoff in the spectrum.  For
example, typical $\gamma$-ray spectra observed by EGRET have a threshold of
100 MeV with spectral indices varying from 1.2 to 2.2 and the cutoffs
vary from 5 to 30 GeV.  For this range, we indicate the average
$\gamma$-ray energies in Figure 2a as a function of both the spectral index
and the cutoff energy.

The contours of average energies in Figure 2a reflect a flat
distribution between 400 to 800 MeV where presumably most of the
$\gamma$-ray pulsars exist. Rather than assuming a constant average energy
for all simulated pulsars, we obtain a rough estimate of how the average
energy varies with the period derivative and period of the pulsar.  The
spectral index of the observed $\gamma$-ray spectrum can be estimated from
the polar cap cascade model (Harding \& Daugherty 1999), which suggests
a dependence upon the parameter $P/B_{12}$.  A crude estimate of this
interdependence can be represented by an expression
\begin{equation}
\alpha =+0.85-0.45\log \left( {{P \over {B_{12}}}} \right).
\label{eq:alph}
\end{equation}
The maximum energy in the $\gamma$-ray spectrum is attributed to the photon
attenuation in the creation of electron-positron pairs as the photon
propagates through the magnetosphere.  The pair creation cutoff energies,
$\epsilon_{max}$, can be derived from cascade codes.  Baring \& Harding
(2000) have formulated the dependence of the maximum energies of photon
escaping the magnetosphere due to pair creation at different altitudes
of emission.  A rough representation of these energies can be made by
the following equation
\begin{equation}
\varepsilon _{\max }({\rm GeV})={{10} \over {B_{12}}}.
\label{eq:emax}
\end{equation}
Having the period derivative and period dependence of the spectral index
and cutoff energy from equations (\ref{eq:alph}) and (\ref{eq:emax}), we
are able to estimate the average energies from equation (\ref{eq:egam})
as a function of the period derivative and period shown in Figure 2b.

\subsection{Emission Geometry}
In this study, we have used a simple geometry for the radio and
$\gamma$-ray emission assuming in both cases a constant emission solid
angle of 1 steradian in which the radio and $\gamma$-ray beams are aligned
and pointing in the direction of the Earth.  There would be $4\pi$ other
pulsars under the same circumstances that would not be detected because
the emission cone would not be in our direction so we do not follow
these in our code. Comparisons are, therefore, made between simulated
radio pulsars and pulsars detected by EGRET and GLAST, observed as point
sources without reference to the photon flux required for the
identification of $\gamma$-ray pulsations.

\subsection{Implementation}
An event is initiated by selecting a neutron star with a random age,
magnetic field and a fixed initial period of 30 ms.  The period is
evolved to the present time assuming dipole spin-down of either a
constant magnetic field or an exponentially decaying magnetic field.
Although the total number of neutron stars are counted, only events to
the left of the multipole death line in the period derivative - period
diagram are further followed.  The neutron star is imparted a kick
velocity due to the supernova explosion, and its trajectory is
integrated from its birth location forward in time to the present.  If
the star is within 30 kpc of the Earth, in the viewing region of one
of the surveys, and a random number is less than the geometric
efficiency of the survey, the simulated pulsar is assigned a radio flux
that is scaled to the observing frequency of the survey as well as a
$\gamma$-ray flux above 100 MeV.  We assume that the radio and
$\gamma$-ray beams are aligned with an effective solid angle of 1
steradian and that the Earth is within the line of sight of the emission
beams. Considering that there are a factor of $4\pi$ other pulsars whose
beams are not in the direction of the Earth, we can estimate a neutron
star birth rate in the Galaxy.  If the scaled radio flux is above the
minimum flux ($S_{\min}$) of the survey, the neutron star is flagged as
a radio pulsar ``observed'' by that survey. Although it may be that
several surveys ``observed'' the same pulsar, we count it as one event. 
Independently from the radio surveys, if the $\gamma$-ray flux is above
the flux thresholds of EGRET or GLAST, it is flagged as a $\gamma$-ray
pulsar observed by either EGRET or GLAST.  In order to obtain smooth
distributions, we continue the process until 10,000 radio pulsars are
observed.  As EGRET viewed (ftp://gamma.gsfc.nasa.gov/pub/PULSAR) each of the observed radio
pulsars in our selected group, we normalize the distributions to the
number of pulsars in the select group of observed radio pulsars. Since
there are 445 observed radio pulsars in our select group that represents
90\% of all the pulsars in the Princeton catalog with the features that
we have specified, this normalization allows for the determination of
neutron star birth rates.  We compare the number of simulated
$\gamma$-ray pulsars predicted to be detected by EGRET with the number
that EGRET actually detected, and also predict the number of
$\gamma$-ray pulsars that GLAST should be able to detect as point and as
pulsed sources.  In this study, we define radio-quiet $\gamma$-ray pulsars as 
those which are detectable by $\gamma$-ray telescopes but not by the radio surveys.
In this manner, we are able to tag the events as
radio-loud and radio-quiet, $\gamma$-ray pulsars ``observed'' by EGRET
and/or GLAST.  Of particular interest is the number of predicted
radio-loud and radio quiet, $\gamma$-ray pulsars, how they compare to
those observed by EGRET as point sources and where each of these source
groups are located in the $\dot P - P$ diagram.

\section{Results}

For the purpose of this study, we have focused on performing simulations
for two different cases: one in which the magnetic field remains constant and
a pulsar death valley in the period derivative-period diagram is assumed and one in
which the magnetic field is allowed to decay exponentially without a
death valley.  We have placed greater emphasis on the case where the
field is constant.

In Figure 3, we present a comparison of the observed (3a) and simulated
(3b) distributions of 445 pulsars shown in Galactic coordinates as
Aitoff projections. The simulation assumes no field decay.  The
distributions are governed in part by the survey regions that have been
chosen, but also by the many primary distributions such as birth
location and Galactic potentials, etc.  As can be seen, the calculated
and observed distributions are very similar.  In this preliminary study,
we do not expect to reproduce the exact numbers observed by each survey.
 The number of pulsars observed depends crucially on the assumptions
made in the determination of $S_{\min}$ and the geometric observing
efficiency.  As mentioned, we have neglected the last  time-smearing
term in equation (\ref{eq:pwid}), we have assumed a constant duty
fraction of 0.05 for the intrinsic pulse width for all periods and all
surveys, we have assumed a constant receiver temperature, and we have
assumed that the efficiency is constant over sky boundaries of the
surveys.  We hope to improve these assumptions in a subsequent study
that includes a realistic representation of the emission geometries for
both the radio and $\gamma$-ray beams.  Yet we affirm that this
preliminary investigation makes important contributions to the
understanding of pulsar emission and paves the way for further studies.

In Figure 4, we compare the period-derivative versus period $(\dot P - P)$
plot of the select group of observed pulsars (4a) with the one of
simulated pulsars (4b). The dotted lines are shown for the locus of
constant magnetic field with the indicated strength.  The dashed lines
represent the indicated ages of pulsars assuming a dipole spin-down
field.  The solid lines show the pulsar death lines for dipole and
multipole magnetic field distributions in the curvature-initiated, space-charge-limited-flow
model (SCLF) (Zhang, Harding \& Muslimov 2000).  The radio pulsars
observed (4a) and those simulated and filtered (4b) through the select group
of surveys are indicated with solid dots.

Radio-loud and radio-quiet, $\gamma$-ray pulsars are represented with solid
triangles and open circles, respectively.  The observed radio-quiet,
$\gamma$-ray pulsar is Geminga and a list of radio-loud, $\gamma$-ray pulsars
detected by EGRET includes Vela, the Crab, B1951+32,
B1706-44, B1509-58, B1055-52, B0656+14 and J1048-5832.  The pulsar
J0218+4232 with a period of 2.32 ms and a period derivative of $7.5
\times 10^{-20}\ {\rm s\cdot s^{-1}}$ is outside the bounds of the plot and of
the characteristics of the select group of pulsars. 
The simulations for this case result in 7 radio-loud and 1 radio-quiet, $\gamma$-ray pulsars ``observed''
by EGRET (shown in Figure 4b) and 76 radio-loud and 74 radio-quiet, 
$\gamma$-ray pulsars ``observed'' by GLAST.   The previously
mentioned death valley operates between these two lines where the code
exponentially ($\propto e^{(-d/\tau)}$) removes excess pulsars with a decay
constant of $\tau=1.0$ in a random fashion depending on the pulsar's
distance, $d$, from the multipole death line along a constant magnetic
field.  Figure 5 shows the simulated distribution using the same set of
parameters as those that generated Figure 4b except without the death
valley in effect.

Clearly the excess of pulsars near the multipole death line in Figure 5
is not seen in the observed distribution in Figure 4a.  However, there
are still significant differences even with the presence of the death
valley in comparing Figures 4a and 4b.  The observed distribution of
radio pulsars is narrower in period, elongated up and down along the
period derivative axis, whereas the simulated distribution is elongated
left and right along the period axis.  The observed distribution has
more pulsars with smaller period derivatives and fewer high field
pulsars in the death valley than simulated distribution.  The observed
distribution of low field pulsars is narrower in period than that
simulated, while the distribution of high field pulsars is broader than
that simulated.  Broadening the primary high field Gaussian (Table 1) or including
a higher field component only increases the number of high field pulsars
bunched up near the multipole death line and cannot reproduce the group
of high field pulsars observed with periods between 0.1 and 2 seconds in
Figure 4a.

The observed group of $\gamma$-ray pulsars is on average younger, with
larger $\dot P$ and higher fields, than the group of simulated pulsars. 
However, the one observed radio-quiet, $\gamma$-ray pulsar is older with a
higher field than the 1 predicted. An important factor in the selection of
the $\gamma$-ray pulsars is the assumed flux
thresholds {\bf (Table 5)}.  
The comparison of the observed and simulated distributions in Figure 4
suggests that there are fewer observed pulsars with higher periods than
in the simulated distribution.  It might be that in addition to the
minimum radio flux dependence on the period described in equation (\ref{eq:smin})
there is a high period limit to the radio surveys where typically the
radio sensitivity increases with increasing period. However, there is in
some published figures a slight increase in $S_{min}$ for pulsars with
periods greater than a few seconds (Dewey et al. 1985 and Stokes et al.
1986), perhaps due to computations of the actual telescope $S_{min}$
during the observing period.  As there are no details given in these
publications, we have not taken this into account in our model simulations.

The clear absence of high-field, high-period observed pulsars in Figure
4a indicated by the shaded circle labeled HB (High field) as compared to
those simulated in Figure 4b, might be suggestive of the decay of the
magnetic field.  There are observed pulsars in the region indicated by
the arrow that should have led to pulsars in the HB region.  In fact,
the whole wedge-shaped distribution of observed pulsars in Figure 4a
could be explained by field decay, including the absence of low-field
observed pulsars in the shaded region labeled LB (low field).  If the
field does not decay, one would expect pulsars in this region (LB) to
have led to the observed low-field pulsars indicated by the arrow.  The excess of
simulated high-field, high-period pulsars have ages of the order of
$10^7$ years.  Therefore, a decay constant of this order is required for
these high-field pulsars to have their fields decay by an order of
magnitude.  In Figure 6, we present a simulation in which we have
included the decay of the magnetic field with a time constant equal to
$5\times 10^6$ years.  In this simulation, we removed the death valley to
facilitate comparison with Figure 5 where we have simulated pulsars
without the pulsar death valley to directly see the effects of the field
decay.  With field decay, the pulsars do not seem to bunch-up along the
multipole death line as in Figure 5 as the field decreases with age,
producing a wedge-shaped distribution at low fields, small period
derivatives and mid-periods.  

We found that the multiple low-field and high-field Gaussians used in the main 
simulations (see
Table 1) were not necessary to achieve a distribution
comparable to the observed distribution in Figure 4a.  The improvement
is quite noticeable.  The wedge-shaped distribution of the observed pulsar 
distribution is explained, even in the region of low period and moderate period
derivative, where the simulation in Figure 4b has an over abundance of
pulsars (LB) in comparison to the observed distribution.  The simulations for this
case result in 9 radio-loud and 2 radio-quiet, $\gamma$-ray pulsars ``observed''
by EGRET (shown in Figure 6) and 90 radio-loud and 101 radio-quiet, $\gamma$-ray
pulsars ``observed'' by GLAST.  Introducing the pulsar death valley along with 
the field decay would remove pulsars from the death valley and yield more pulsars
to the left of the valley, thereby, increasing the number of radio-loud, $\gamma$-ray pulsars.
The simulation with field decay suggests that the death valley might not be a required
artifact to have reasonable agreement with the number of radio-loud, and radio-quiet, 
$\gamma$-ray pulsars detected by EGRET.  The main difficulty
in accounting for the observed shape of the pulsar distribution by field
decay is being able to justify the small time constant of $5\times 10^6$ 
years.  Several mechanisms for field decay in
neutron stars have been studied (e.g. Goldreich \& Reisenegger 1992).
Ohmic diffusion, with a decay timescale $\sim 2 \times 10^{11}{\rm\ yr}\,
L_5^2/T_8^2$, where $L_5$ is the characteristic length scale of the flux 
loops in units of $10^5$ cm and $T_8$ is the core temperature in units of
$10^8$ K, dominates in fields $B < 10^{11}$ G.  Hall drift 
in the crust, with a timescale $\sim 5 \times 10^{8}{\rm\ yr}\,
L_5^2 T_8^2/B_{12}$, dominates in fields $B \sim 10^{12} - 10^{13}$
G.  Ambipolar diffusion, with a decay timescale $\sim 3 \times 10^{9}
{\rm\ yr}\,L_5^2 T_8^2 / B_{12}^2$, dominates at the highest fields.
Thus, for fields $B \sim 10^9 - 10^{13}$ characteristic of the radio 
pulsar population, the decay timescale is longer than $3 \times 10^7$ yr.
However, if the birth distribution of fields has a mean of around
$2 \times 10^{13}$ G, then the average initial 
decay timescale would be $\sim 5 \times 10^6$ yr.  But since the decay
timescale is a strong function of field strength, this initially short
decay timescale would not persist as the field decreases.  Clearly, a
more detailed study is required to fully address the effect of realistic
field decay in the radio pulsar population studies.

In Figure 7, we compare various distributions of the indicated features
of observed pulsars (shaded) and of simulated pulsars.  The smooth
simulated histograms have been obtained from a group of 10,000 pulsars
and then normalized to a total of 445 pulsars, the number of the
selected group of observed pulsars. The dark histograms represent the
case in which there is no field decay and a pulsar death valley is
assumed as discussed previously.  The light histograms result from the
simulation of the field decay case with no death valley.  Under the
assumption of field decay, the pulsar age depends on the decay constant
of the magnetic field (5 Myr).  As a result, we show separate figures
for the pulsar age distributions.  As noted for the case of no field
decay, too many older pulsars with long periods and with small period
derivatives are produced in our simulation.  However, the simulated
distributions of radio flux and distance from Earth along with the
magnetic field agree very well with those observed.  The distributions
for the case of field decay overall appear to better describe the
observed distributions without the necessity of introducing a pulsar
death valley between the dipole and multipole death lines.

The period and period derivative dependence of the radio luminosity is
another important function that crucially determines the radio selection
of simulated pulsars.  As mentioned previously, we have used the
functional form developed by Bhattachary et al. (1992) who followed the
work of Narayan \& Ostriker (1990), Vivekanand \& Narayan (1981) and Pr\'{o}szy\'{n}ski \&
Przybyci\'{e}n (1984) and found the best fit for a luminosity function
described by equation (\ref{eq:l400}).  While we do not wish to develop a fitting
model in this study, we found that it was necessary to slightly adjust
the intercept of their functional form from 6.64 to 7.2.  We present in
Figure 8 a comparison of the radio luminosity distribution at 400 MHz
$(\log_{10}\langle L_{400}\rangle)$ as a function of $\log_{10}(\dot{P}/P^3)$
and the population distributions of the radio luminosity for observed
and simulated pulsars displayed as histograms.  In the
$\log_{10}(L_{400})$ versus $\log_{10}(\dot P/P^3)$ plots, we performed
linear fits in order to make comparisons of the distributions.  The
resulting fit parameters are indicated in Table 6.

\begin{center}
\begin{tabular}{ccc}
\multicolumn{3}{c}{Table 6} \\ \hline
\multicolumn{3}{c}{Linear Fit Parameters} \\ \hline
	& Intercept	& Slope \\ \hline
Observed	& $5.2 \pm 0.4$	& $0.21 \pm 0.03$ \\
6.64 intercept in Eqn. 22	& $5.0 \pm 0.3$	& $0.23 \pm 0.02$ \\ 
7.2 intercept in Eqn. 22	& $5.2 \pm 0.3$	& $0.21 \pm 0.02$ \\ \hline
\end{tabular}
\end{center}

Note that we do not expect to obtain the same parameters as in equation
(\ref{eq:l400}), which represents the average radio luminosity of the 
primary pulsars
prior to being observed. {\bf The modified luminosity law with an intercept of 7.2
also fits well the luminosity distributions for the case assuming the decay of the magnetic field.}
The actual radio luminosity at 400 MHz is
dithered about the average using the function described in equation (\ref{eq:dither})
and (\ref{eq:lamb}).  Hartman et al. (1997) describe how the parameters of the
dithering function affect the net luminosity distribution.  The work of
Narayan \& Ostriker (1990) was done with a sample of 265 pulsars known
at the time with measured period and period derivative obtaining a
functional form that has been used quite often in the literature.
However, it seems clear that with a much larger sample of pulsars, a
better functional form needs to be obtained.  This functional form of
the radio luminosity plays an important role in the selection of a radio
pulsar and in the ratio of radio-quiet to radio-loud, $\gamma$-ray pulsars.
 Clearly the required dithering about the average luminosity masks some
dependence that is currently not understood, perhaps reflecting some
dependence of the emission geometry on the period as suggested
by Kijak \& Gil (1998) and Rankin (1993).

Another important parameterization of the primary pulsars is the initial
velocity kicks given to pulsars at the time of their birth from
asymmetric supernova explosions.  These velocity distributions have been
discussed extensively by Dewey \& Cordes (1987), Lyne \& Lorimer (1994),
Bailes (1989), van den Heuvel \& van Paradijs (1997), and Lorimer,
Bailes \& Harrison (1997) indicating space birth velocities of 400 -
500 km/s.  Hansen \& Phinney (1997) indicate that these studies did not
include the selection effects from flux limits of the pulsar surveys or
the accuracy of the proper motion determinations and conclude that a
Maxwellian distribution with a mean velocity of 300 km/s adequately
describes the observations.  More recently, a study undertaken by Cordes
\& Chernoff (1998) concludes that the three-dimensional velocity
distributions can be best accounted for using a two component Gaussian
function with mean velocities of 175 and 700 km/s representing 86 \%
and 14 \% of the population.  In this study, we have chosen to use the
functional form of Lyne \& Lorimer (1994) with some modification.  We
used the $z$ distribution of the radio pulsars from the Galactic disk to
establish the kick velocity parameter.  In Figure 9, we present the $z$
distribution (shaded) of observed pulsars from our select group of 445.

The spikes at $\approx 1.7$ kpc reflect the limitation of the distance
model (Taylor \& Cordes 1993).  The thin and thick  solid histograms
represent the simulated distributions using $\zeta=v/120$ km/s used in this
study and $\zeta=v/350$ km/s used by Lyne \& Lorimer (1994), respectively. 
Since the primary $z$ distribution of pulsars is assumed to be
exponential, we have fit the observed and simulated distributions shown
in Figure 9 with exponential forms obtaining the following widths shown
in Table 7.

\begin{center}
\begin{tabular}{ccc}
\multicolumn{3}{c}{Table 7} \\ \hline
\multicolumn{3}{c}{Widths (kpc) of Exponential Fits of the z Distribution} \\
Observed & For $\zeta=v/120\ {\rm km/s}$ & For $\zeta=v/350\ {\rm km/s}$ \\ \hline
$0.18\pm 0.01$ &	$0.18\pm 0.01$ &	$0.34\pm 0.01$ \\ \hline
\end{tabular}
\end{center}

The functional form of Lyne \& Lorimer with a parameter of $\zeta=v/350$
km/s results in a distribution with most probable and average
velocities of 266 and 494 km/s, respectively, and leads to a
significantly broader distribution twice the width of the observed
distribution.  Much better agreement is obtained with a parameter of
$\zeta=v/120$ km/s with a most probable and average velocities of 91 and
$168$ km/s, respectively.  We do find that the $z$ distribution is
slightly sensitive to other parameters of our model described in this
study.  Again we are trying to use the best available distributions of
pulsar features, but comparing the resulting distributions ``by eye'' has
required us to make some small changes.  {\bf The agreement is equally good 
for the case assuming the decay of the magnetic field with a slight narrower 
distribution with a width of $0.14 \pm 0.01\ kpc$.  As a result, we have
used this same velocity distribution for both the cases where the field is constant
and is allowed to decay.}

In Figure 10, we present a set of $\dot P - P$ diagrams for radio-quiet,
$\gamma$-ray pulsars as seen by EGRET (10a) and GLAST (10b) and for radio
loud, $\gamma$-ray pulsars as seen by EGRET (10c) and GLAST (10d) represented
by solid circles.  We have also included the group of $\gamma$-ray pulsars
detected by EGRET in Figures 10a and 10c shown by solid triangles.  In
order to obtain smoother distributions of these different groups in the
$\dot P P$ diagram, we have simulated a group of 10,000 pulsars and
depict their distributions by the contoured regions.  The gray scaled
background represents the energy-integrated $\gamma$-ray and average radio
luminosities in the upper panels (10a and 10b) and in the lower panels (10c
and 10d), respectively.  The average radio luminosity is proportional to
$\dot P^{1/3} P^{-1}$, while the $\gamma$-ray luminosity in regime II
drops faster and is proportional to $\dot P^{1/2} P^{-7/4}$.  The most
intense luminosities for both radio and $\gamma$ rays are in the upper left
in the $\dot P - P$ diagram, but here the pulsars are very young and few
in number.  The numbers indicated in the legend represent the number of
simulated pulsars from the group of 445 radio pulsars.  The model
predicts that GLAST should observe 76 radio-loud, $\gamma$-ray pulsars
compared to 7 predicted for EGRET detected as point sources.  The model
predicts that GLAST should observe 74 radio-quiet, $\gamma$-ray pulsars
compared to 1 predicted for EGRET.  The GLAST sensitivity for blind period
searches is expected to be about the same as the EGRET point-source
detection sensitivity (S. Ritz, priv. comm.).  GLAST will, therefore, be
expected to identify 7 of these 74 objects as pulsed sources.
The distribution of radio-loud, $\gamma$-ray pulsars predicted to be
observed by GLAST peaks at a lower $\gamma$-ray luminosity towards the more
populated region of radio pulsars than the distribution predicted for
EGRET.  Due to the greater sensitivity of GLAST, the group of radio
quiet, $\gamma$-ray pulsars also moves toward lower $\gamma$-ray luminosities
and more populated regions in the $\dot P - P$ space, but the radio-quiet pulsars
are younger.

The radio-quiet pulsar, Geminga, observed by EGRET is located in the
GLAST region of radio-quiet pulsars in Fig. 10, but somewhat removed from 
the EGRET region.  The results of the simulation seem to agree fairly well with
the observations made by EGRET though, perhaps, predicting a few too
many radio-loud, $\gamma$-ray pulsars.  As mentioned before, many of the
EGRET unidentified sources are expected to be radio-quiet, $\gamma$-ray
pulsars.

The results predicted for GLAST are interesting in that about as many
radio-quiet, $\gamma$-ray pulsars are expected to be observed as
radio-loud, $\gamma$-ray pulsars. Given the flux thresholds we have used,
these are objects observed as point sources.  In order to understand why
GLAST is predicted to detect a larger ratio of radio-quiet to radio-loud, 
$\gamma$-ray pulsars than EGRET, we present in
Figure 11 the distribution of the distance of the pulsar from Earth for
each of these groups. We simulated a group of 10,000 radio pulsars and
normalized the distributions to the number (445) of observed radio
pulsars in the selected group.  As a result, one can see fractional
numbers of pulsars in the distributions.  The distribution of observed
radio pulsars is presented as a shaded histogram.  The
predicted distributions of radio-quiet (thin) and radio-loud (thick),
$\gamma$-ray pulsars for GLAST are displayed by the plain histograms.  The
predicted distributions of radio-quiet (diagonal pattern) and radio-loud
(brick pattern), $\gamma$-ray pulsars are indicated for EGRET.  The
distribution of radio-loud, $\gamma$-ray pulsars detected by both EGRET and
GLAST have distributions with similar shapes as the observed radio distribution. 
Due to significantly different $\gamma$-ray flux thresholds, the
distributions of radio-quiet, $\gamma$-ray pulsars are quite different for
EGRET and GLAST.  Thus the simulation suggests that due to the increased
sensitivity, GLAST will be able to detect more pulsars that are further
away than even those detected by the radio surveys used in this study.  This 
may change with the observations from the Parkes multibeam pulsar survey 
(Manchester et al. 2001) that has been finding more young, distant pulsars.

In Table 8, we summarize the number of radio-quiet and radio-loud, $\gamma$-ray pulsars
simulated for the two main cases explored in this study in which we have assumed
no magnetic field decay requiring a pulsar death valley and field decay without a death valley.
The results for the case of no field decay without a valley are also included for 
comparison.  In addition, we have indicated the number of pulsed sources that GLAST would
observe from the radio-quiet, $\gamma$-ray pulsar group.  

\begin{center}
\begin{tabular}{c|c|c|c|c|c}
\multicolumn{6}{c}{Table 8} \\ \hline
\multicolumn{6}{c}{Simulated Pulsar Statistics} \\ \hline\hline
 & \multicolumn{2}{|c|}{EGRET} & \multicolumn{2}{|c|}{GLAST} & Neutron Star Birth \\
Case & Radio Quiet & Radio Loud & Radio Quiet (Pulsed) & Radio Loud & Rate (per century) \\ \hline
No decay  & & & & & \\ 
with valley & 1 & 7 & 74 (7) & 76 & 1 \\ \hline
No decay & & & & & \\
no valley & 1 & 3 & 35 (3) & 38 & 0.5 \\ \hline
Decay & & & & & \\
no valley & 2 & 9 & 101 (9) & 90 & 2 \\ \hline 
\end{tabular}
\end{center}
These simulations are able to suggest a neutron star birth rate.  As
mentioned previously, we have assumed a 1 steradian solid angle of the
radio beam, so there is a factor $4\pi$ more pulsars whose beams do not
point in the direction of the Earth than those used in the simulation.
We indicate in the table the estimated neutron star birth rates per
century for each of the indicated cases.  We have corrected the birth
rates assuming a Gaussian beam to correct the overall detection by the
ratio of the detection volume to that of the actual volume that
increases the birth rate by a factor of 1.4 (Arzoumanian, Chernoff \&
Cordes 2001).

\section{Discussion}

We have modeled the Galactic population of radio pulsars using a
Monte Carlo simulation in order to calculate the expected numbers
of $\gamma$-ray pulsars detectable by EGRET and GLAST.  This paper
has focused on the $\gamma$-ray luminosity predicted by the polar
cap model of Daugherty \& Harding (1996) and Zhang \& Harding (2000).
Our simulation predicts about the same number of radio-loud
$\gamma$-ray pulsars (7) compared to the number (8) detected by EGRET. 
We predict that GLAST should detect on the order of {bf 75-100} radio-loud pulsars,
however this number could also change with a detailed treatment of
geometry.  A very interesting, and somewhat unexpected, result of our
simulation is the prediction that GLAST will detect about the same number (74) 
of radio-quiet as radio-loud pulsars (76), at least as point sources.  Since
we have assumed that the radio and $\gamma$-ray beams are identical in
this study, this result implies that GLAST will be more sensitive than
radio surveys with regard to the detection
of young pulsars in the Galaxy.  Even with the considerable model 
uncertainties, we can conclude that polar cap models predict a much smaller
ratio of radio-quiet to radio-loud $\gamma$-ray pulsars than do outer gap 
model studies. For example, Zhang, Zhang \& Cheng  (2000) predict that GLAST will detect 
only about 80 radio-loud $\gamma$-ray pulsars but about 1100 radio-quiet 
$\gamma$-ray pulsars.
This ratio then, regardless of the exact numbers, will be an important 
discriminator between polar cap and outer gap models.

We have found that magnetic field decay could have a significant effect
on pulsar evolution and the formation of the observed pulsar $\dot P-P$
distribution. As mentioned previously, there are distinct differences 
between the observed distribution of Figure 4a and the simulation without 
field decay of Figure 4b.  There is an over
abundance of simulated pulsars with high fields and high periods.  The
apparent drop in observed pulsars in this region could be conceived as
evidence for the decay of the magnetic field.  Our preliminary
calculations suggest that a decay constant of the order of $10^7$ years
is required to effect such a distribution, as also suggested by the ages
of these pulsars in this region.  A recent study of the radio pulsar
distribution by Tauris \& Konar (2001) also concludes that some torque decay
is required.  Most studies of field decay estimate
longer decay constants in fields below $B \sim 10^{13}$ G.  However,
we believe that this issue warrants further study. 

We can analytically calculate the time it takes a pulsar from birth with
a given magnetic field to reach the multipole death line.  Since we are
assuming a constant birth rate, dividing this time by the maximum age of
$10^9$ years used in the simulation provides a survival fraction of
initial pulsars found to the left of the multipole line and is given by the expression
\begin{equation}
F_s=9.0\times 10^{-3}B_{12}^{-6/7}.
\end{equation}
From this equation, one can see that as the magnetic field increases 
by an order of magnitude, the survival
fraction will decrease by an order of magnitude.  This
explains why
the factor $A_3$ in Table 1, though much smaller than $A_1$, makes a
significant contribution.  The initial population is weighted by this
primary distribution of the magnetic field.  Looking at the wings of the
field distribution say at $10^{13}$ G and $10^{11}$ G, assuming no
selection effects, these pulsars would be randomly distributed over time
with 90\% of them lying between the age lines of $10^8$ and $10^9$ years.  However,
the radio luminosity, being proportional to $\dot P^{1/3} P^{-1}$,  is
dropping toward the multipole line.  Therefore, pulsars with shorter
periods to the left of the multipole line are more radio luminous and 
are seen scattered to the left of the line for an age of
$10^8$ years. The simulation indicates a broader distribution in period
along the $B=10^{11}$ G line than the distribution along the $B=10^{13}$
G line.  The observed distribution shows the opposite effect in these
regions with a narrower distribution along the $B=10^{11}$ G line with
very few pulsars with periods less than 0.2 seconds and a broader
distribution along the line $B=10^{13}$ G that cuts off the number of pulsars dramatically
beyond a 2 second period.  The paucity of pulsars at low periods is
maybe related to several effects that we have not accounted for in this simulation.
Perhaps we have underestimated the actual minimum radio flux
threshold given theoretically by equation (\ref{eq:smin}) and plotted in Figure 1. 
The flux threshold tends to increase dramatically for some surveys for
periods less than 0.1 s (as in the cases the Molonglo 2 and Green Bank 2
surveys with rather large sampling rates), which is where the observed distribution in Figure 4a begins
to indicate fewer pulsars than the simulated distribution in Figure 4b.
This may be a result of the fact that we have left out the fifth
term in equation (\ref{eq:pwid}), which becomes significant for low period pulsars
and is more significant for the Green Bank 3 and Arecibo 2 surveys.  We
hope to add this term to our next simulations when we also take into 
account the emission geometry.  However, this is not a problem for the newer surveys
with smaller sampling rates.  With the exception of the Jodrell Bank 2 and Parkes 1 survey,
the surveys were performed at low frequencies around 420 MHz.  Young pulsars
with short periods and large period derivatives are distant and in the Galactic
plane where the effects of scattering and background temperatures
require a more careful treatment than we have done in this simulation.

On the other hand, the group of high field, large period pulsars missing
in the observed distribution may be a manifestation of the effects of
the geometry of the radio beam.  Various studies (Arzoumanian, Chernoff \& Cordes
2001, Kijak \& Gil 1998, Gil
\& Han 1996 and Rankin 1993) suggest that the pulse width of both the
core and conal emission beams is proportional to $P^{-1/2}$.  Therefore,
as the period increases, the solid angle of the emission decreases.  The
observed distribution in Figure 4a seems to indicate a decrease of
pulsars above a period of about 2 seconds across magnetic field
strengths higher than $10^{12}$ G to the left of the multipole death
line, which is also reflected in Figure 7 where the simulation predicts
too many older pulsars. We expect that our forthcoming model with more realistic
geometries will yield better agreement in this region.

In our model, we introduced a death valley to drastically reduce the
number of pulsars between the dipole death lines and the multipole death
line.  As the pulsars age along constant field lines between these lines
both the $\gamma$-ray and radio luminosities are decreasing as well as the
spin-down energy.  Cascade simulations (Baring \& Harding
2001) suggest that the density of electron-positron pairs is also
decreasing in a similar fashion.  While there has been no firm
theoretical development to connect the pair density and the radio
luminosity, one can speculate that there must be a definite relationship,
and the radio luminosity must decrease in some manner with decreasing
pair density. Perhaps the need for the death valley also reflects the
geometric relationship of the radio beam to the period, as discussed above.  

There is also an interdependence of the
derived velocity distribution, radio luminosity function and field evolution. 
For example, the velocity distribution predicted by Lyne \&
Lorimer (1994) does not fit the observed z distribution assuming the luminosity
model of Narayan \& Ostriker (1990).  Using their
distribution, we find a width of the z distribution that is twice that
of the observed distribution as shown in Figure 9.  Our simulations
suggest that most of the pulsars can be accounted for with a smaller
mean velocity of 170 km/s, {\bf which we found to be essentially independent on
whether the field decays or not.}  Clearly there are pulsars with space
velocities of at least one thousand km/s.  However, we did not attempt to
introduce a second distribution of high velocities as in the case of the
study by Cordes \& Chernoff (1998) and Arzoumanian, Chernoff \& Cordes
(2001).  We have also been careful to check that the distance distribution
of the simulated  pulsars agrees with the observed distribution as noted in
Figure 7.  If the pulsars are too radio bright, more pulsars further away will 
be detected in the simulation causing the distribution to shift further away.  
Assuming the distance model is correct, we adjusted the 
Narayan \& Ostriker (1990) luminosity law based on the observed luminosity distributions
shown in Figure 8.  The luminosity model of Narayan \& Ostriker (1990) is
also somewhat dependent on their assumption of field decay, for which they  
studied several cases with decay constants ranging from 8 million to 11 million years.
Bhattacharya et al. (1992), using the same luminosity model, found similar results 
for decay constants of 10 and 100 million years, although the study of Hartman 
et al. (1997), which did not include magnetic field decay, also used this same 
luminosity law.  So the luminosity model, velocity distribution and field
evolution remain uncertain, especially  without the inclusion of the geometry of 
the radio beams.

A population of radio-quiet, $\gamma$-ray pulsars
results very naturally from the different dependence upon the pulsar
period and period derivative of the radio and $\gamma$-ray luminosities. 
With the same beam geometry for both radio and $\gamma$-ray emission, the
model predicts a few radio-quiet, $\gamma$-ray pulsars, like Geminga, that
EGRET should have observed as well as about the right number of
radio-loud, $\gamma$-ray pulsars.  The number if radio-quiet $\gamma$-ray 
pulsars is likely to increase
with a more realistic treatment of beaming geometry. 
Within the EGRET Third Catalog, there
are 171 unidentified $\gamma$-ray point sources, many of which could be
radio-quiet, $\gamma$-ray pulsars, an issue that will be settled by GLAST
with its ability to perform period searches.  

If there were no selection effects imposed, the main concentration of
pulsars would be near the multipole death line.  Since the average radio
luminosity is proportional to $\dot P^{1/3} P^{-1}$, the observed radio
population is pushed away from the multipole death line towards the
upper left portion of the $\dot P - P$ diagram.  The average $\gamma$-ray luminosity
is proportional to $\dot P^{1/2} P^{-7/4}$  and, therefore, drops faster
than the radio luminosity pushing the population of $\gamma$-ray pulsars
even higher toward the upper left, especially in the case of EGRET with
higher thresholds than GLAST as seen in Figure 10.  With greater
sensitivities of GLAST, the $\gamma$-ray pulsar population moves more to
the lower right to correspond more with the region of the population of
radio pulsars as also noted in Figure 10d. However, with the lower
$\gamma$-ray flux thresholds expected for GLAST, it will be able to observe
more pulsars at larger distances than those observed by the radio
telescopes that were used in the observations of the select group of
pulsars in the catalog.  As a result, the expected number of
radio-quiet, $\gamma$-ray pulsars increases relative to the radio-loud,
$\gamma$-ray pulsars from those observed by EGRET.  

The ratio of
radio-quiet to radio-loud, $\gamma$-ray pulsars is strongly sensitive to
the assumptions for the geometries of the radio and $\gamma$-ray beams. In
this study, we assumed a simple geometry that accords with the polar cap
model where the region of radio and $\gamma$-ray emission are tied to the
magnetic polar cap area and are very similar.  The outer gap model, on
the other hand, suggests a very different geometry for the $\gamma$-ray
beam, which originates on field lines of the opposite pole from that
of the visible radio beam.  These outer
gap models (Romani \& Yadigaroglu 1995, Cheng \& Zhang 1998, Zhang, Zhang 
\& Cheng 2000) predict
that GLAST will detect many more radio-quiet, $\gamma$-ray pulsars.  
In a model independent
study, MacLaughlin \& Cordes (2001) simply assume a large solid angle ($2\pi$) 
for the $\gamma$-ray beam and predict that GLAST will
detect around 120 radio-loud and 750 radio-quiet pulsars.
Polar cap models predict that older pulsars with ages greater than $10^6$
years and with longer periods should emit $\gamma$ rays, while the outer
gap models predict that such pulsars will be radio-quiet (Chen \&
Ruderman 1993).  {bf While the simple model studied here} provides strong diagnostics, in
the near future, we hope to perform simulations with a model that
includes a more realistic geometry for the radio as well as the
$\gamma$-ray beams and should provide further understanding that may help
differentiate between pulsar models.

\acknowledgments

We would like to express our thanks and appreciation to Zaven Arzoumanian
at Goddard for his clear insight and assistance in interpreting the
parameters associated with the different radio surveys, to Steve Ritz
and Dave Thompson for discussion of EGRET and GLAST detection properties
and to Bing Zhang and Matthew Baring for helpful comments on the manuscript.  
In addition, we are grateful for the many constructive comments offered to us by
the referee of the paper.
We also express
our sincere appreciation for the generous support of the Michigan Space
Grant Consortium, of the Research Corporation, of the National
Science Foundation under the REU program and through the grant
NSF-9876670, and of the NASA Astrophysics Theory Program.


\newpage

\begin{figure}
\epsscale{0.5} 
\plotone{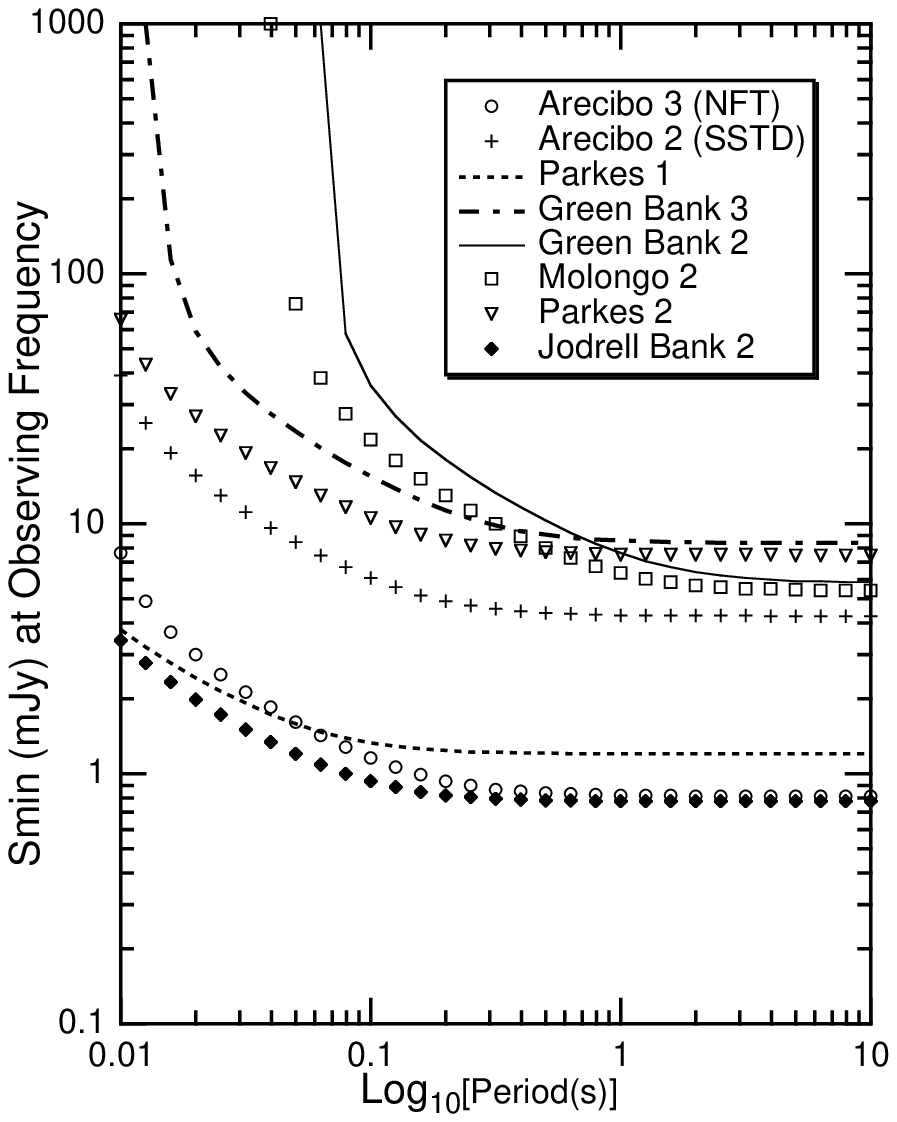} 
\caption{Radio flux thresholds, $S_{min}$, at the observing frequency for each of the
eight surveys used in the simulation as a function of the pulsar period assuming
a sky temperature of 150 K at 408 MHz and a dispersion measure of 200 ${\rm
cm^{-3}\cdot pc}$.}
\end{figure}

\begin{figure}
\epsscale{1.0} 
\plotone{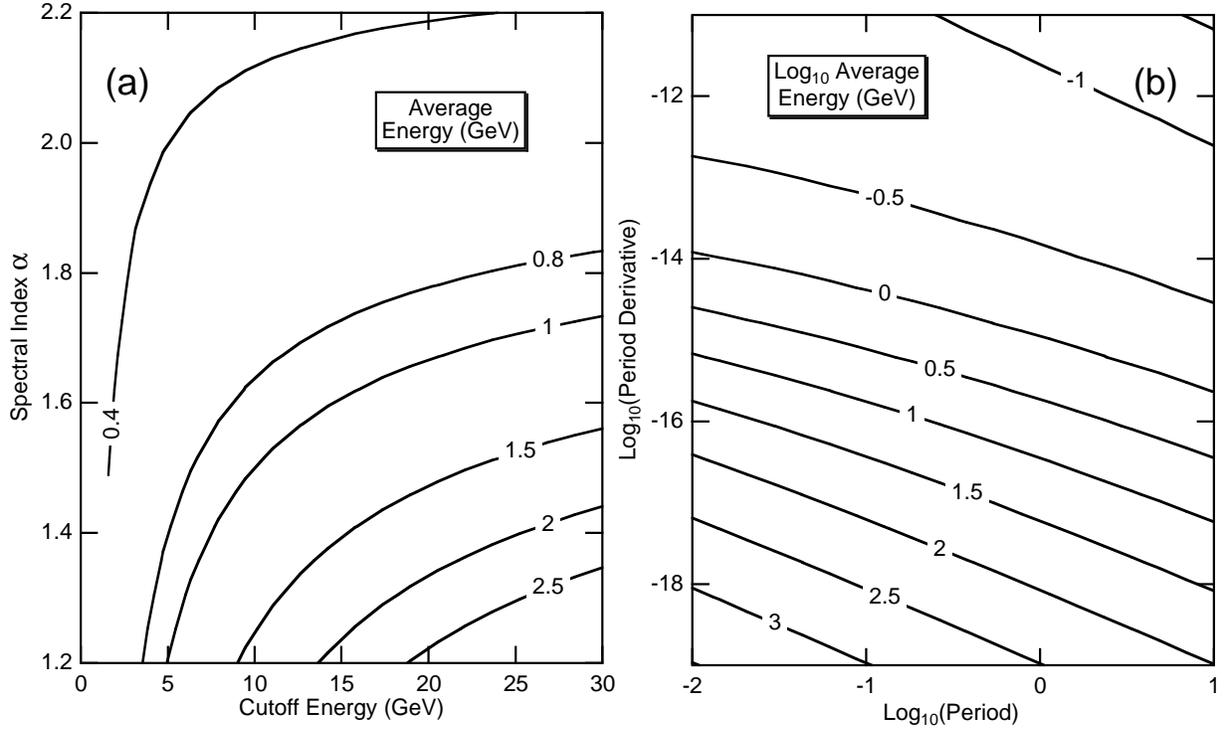} 
\caption{The average $\gamma$-ray energy (contours in GeV) from equation (\ref{eq:egam}) is
plotted (a) as a function of spectral index and  the high energy cutoff assuming
a threshold energy, $\epsilon_{th}$ = 100 MeV, typical for EGRET.  {bf (b)} 
The logarithm of the
average energy (contours) in GeV as a function of the logarithms of the period
derivative and the period from equations (\ref{eq:egam}) to (\ref{eq:emax}).}
\end{figure}
\begin{figure}
\epsscale{1.0} 
\plotone{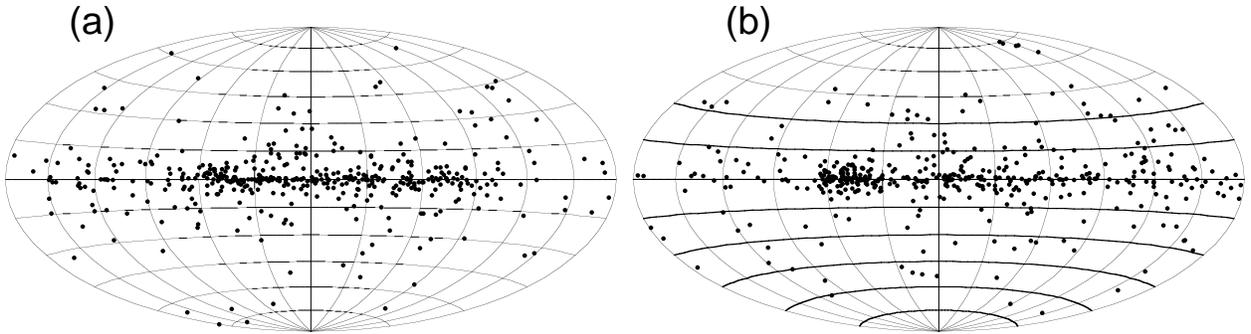} 
\caption{Aitoff plots of the observed pulsars (a) and of the simulated pulsars (b) for the case
of no field decay and a death valley.}
\end{figure}
\begin{figure}
\epsscale{1.0} 
\plotone{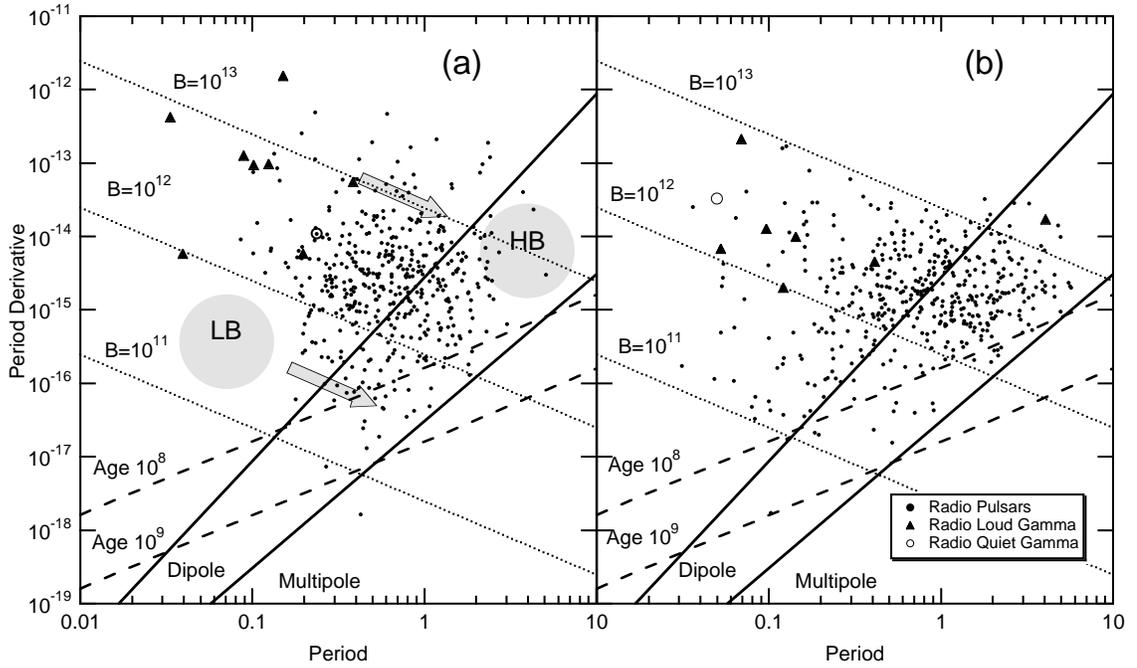} 
\caption{Distributions of observed pulsars (a) and simulated pulsars (b) for the case of
no field decay and a death valley as a
function of the period derivative and period (in seconds) of the pulsars.  Solid dots indicate
radio pulsars. Solid triangles represent radio-loud, $\gamma$-ray pulsars and open
circles symbolize radio-quiet, $\gamma$-ray pulsars observed (a) and predicted (b)
for EGRET.}
\end{figure}
\begin{figure}
\epsscale{1.0} 
\plotone{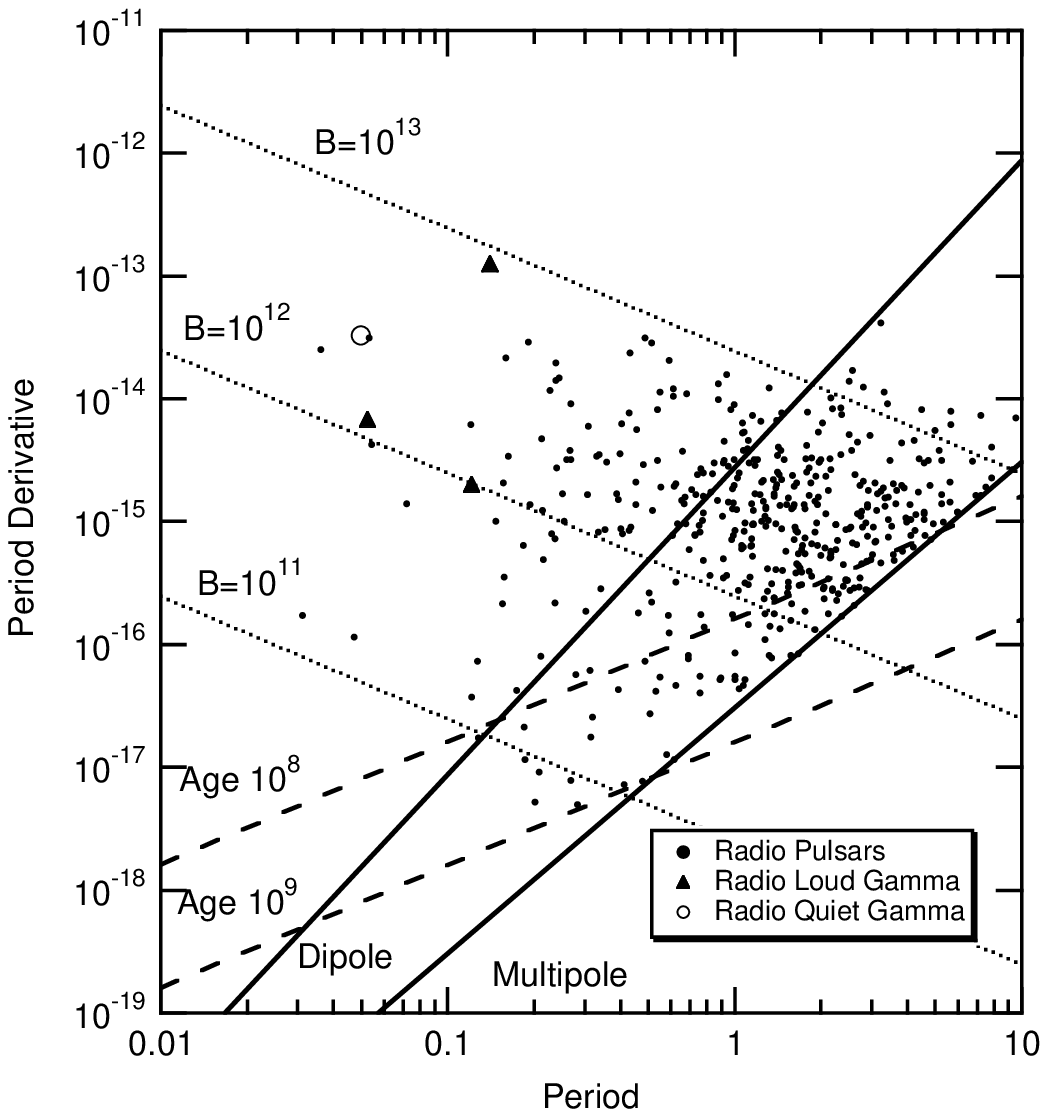} 
\caption{Distributions of simulated pulsars as a function of the period
derivative and period for case of no field decay and no death valley.  Solid
dots indicate radio pulsars.  Solid triangles represent radio-loud, $\gamma$-ray
pulsars and open circles symbolize radio-quiet, $\gamma$-ray pulsars predicted for
EGRET.}
\end{figure}
\begin{figure}
\epsscale{1.0} 
\plotone{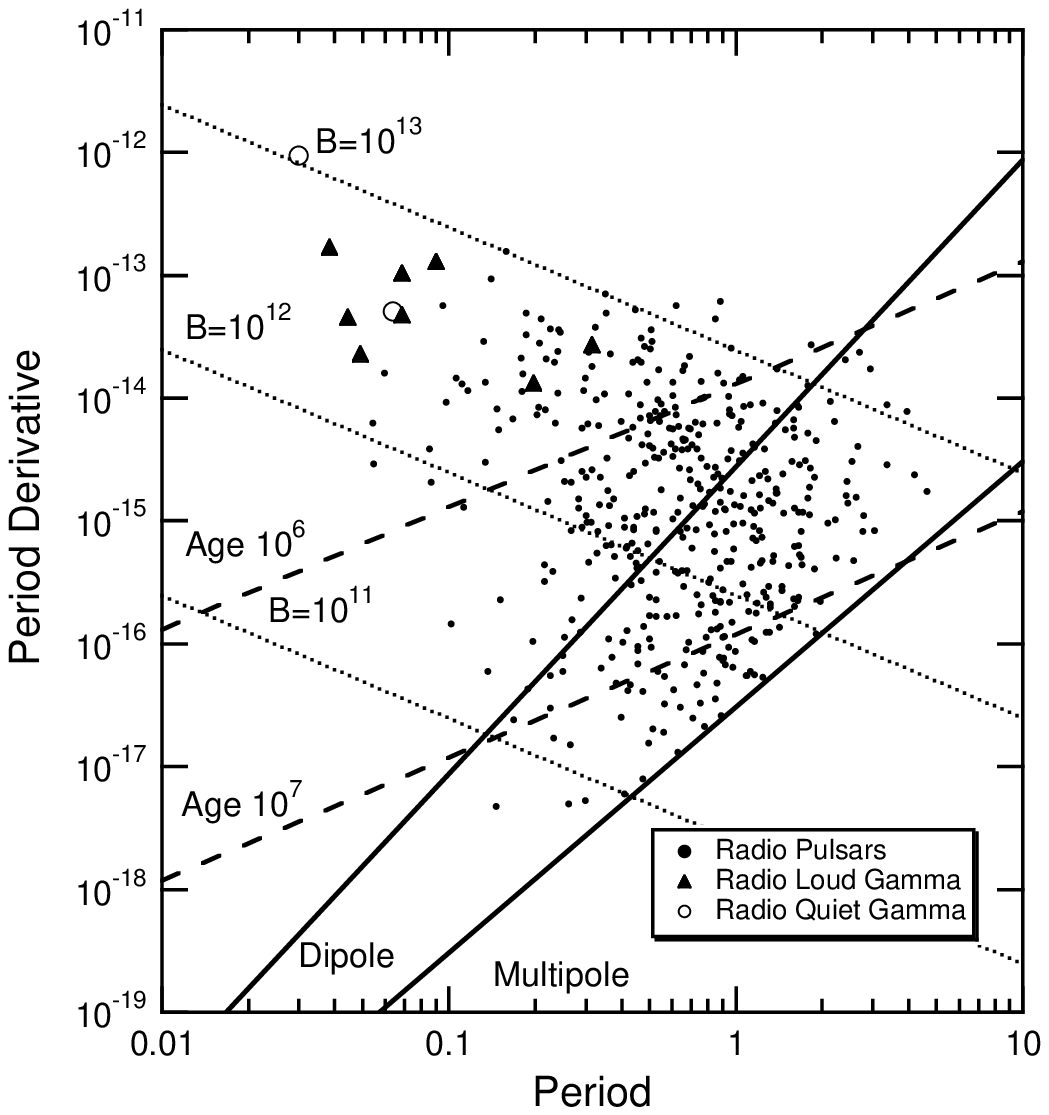} 
\caption{Distributions of simulated pulsars as a function of the period
derivative and period of simulated pulsars for case of magnetic field decay, assuming a time constant of $5\times 10^6$ years, and no pulsar death valley.  Solid
dots indicate radio pulsars.  Solid triangles represent radio-loud, $\gamma$-ray
pulsars and open circles symbolize radio-quiet, $\gamma$-ray pulsars predicted for
EGRET.}
\end{figure}
\begin{figure}
\epsscale{0.75} 
\plotone{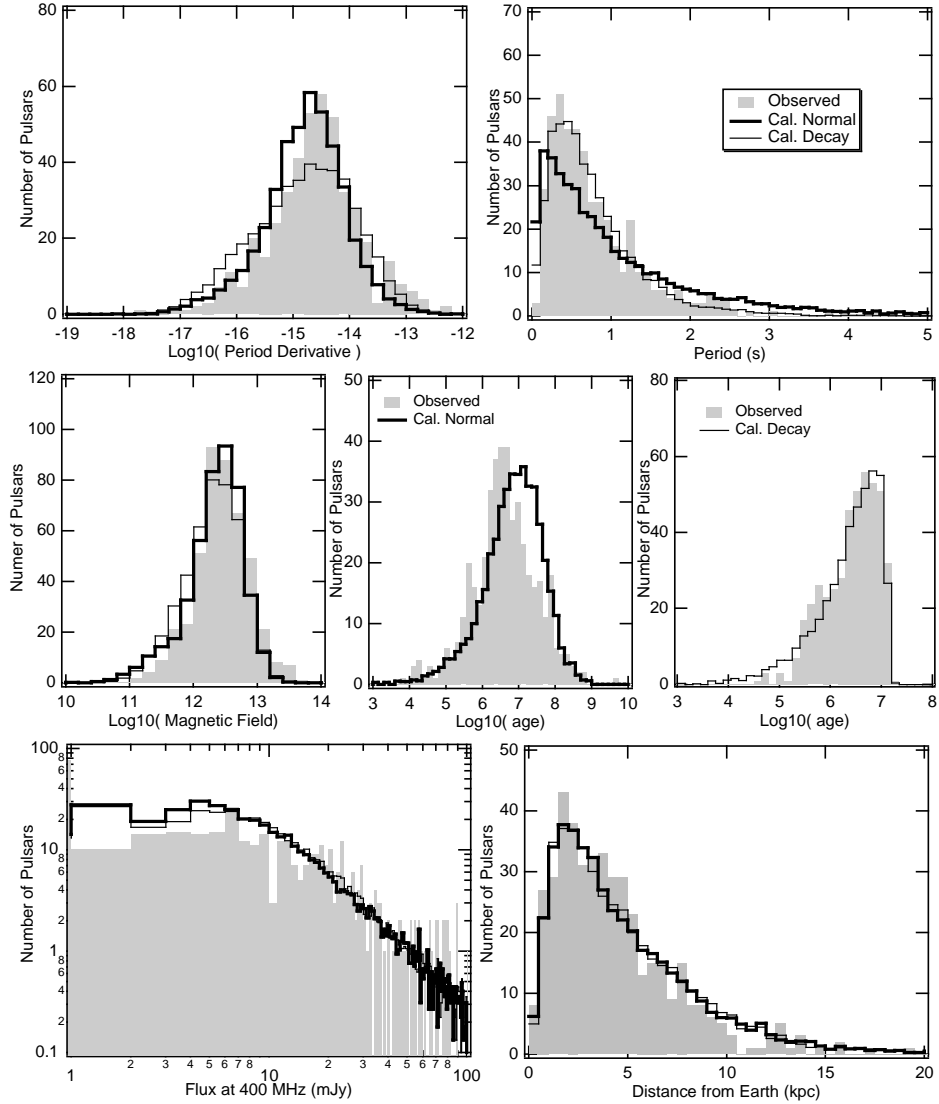} 
\caption{Distributions of various pulsar characteristics indicated as shaded
histograms (observed pulsars) and plain histograms (simulated pulsars).  Thick
histograms represent the distributions of the simulated
pulsars assuming no field decay and pulsar death valley, while the thin
histograms result from the case assuming field decay and no death valley.}
\end{figure}
\begin{figure}
\epsscale{0.75} 
\plotone{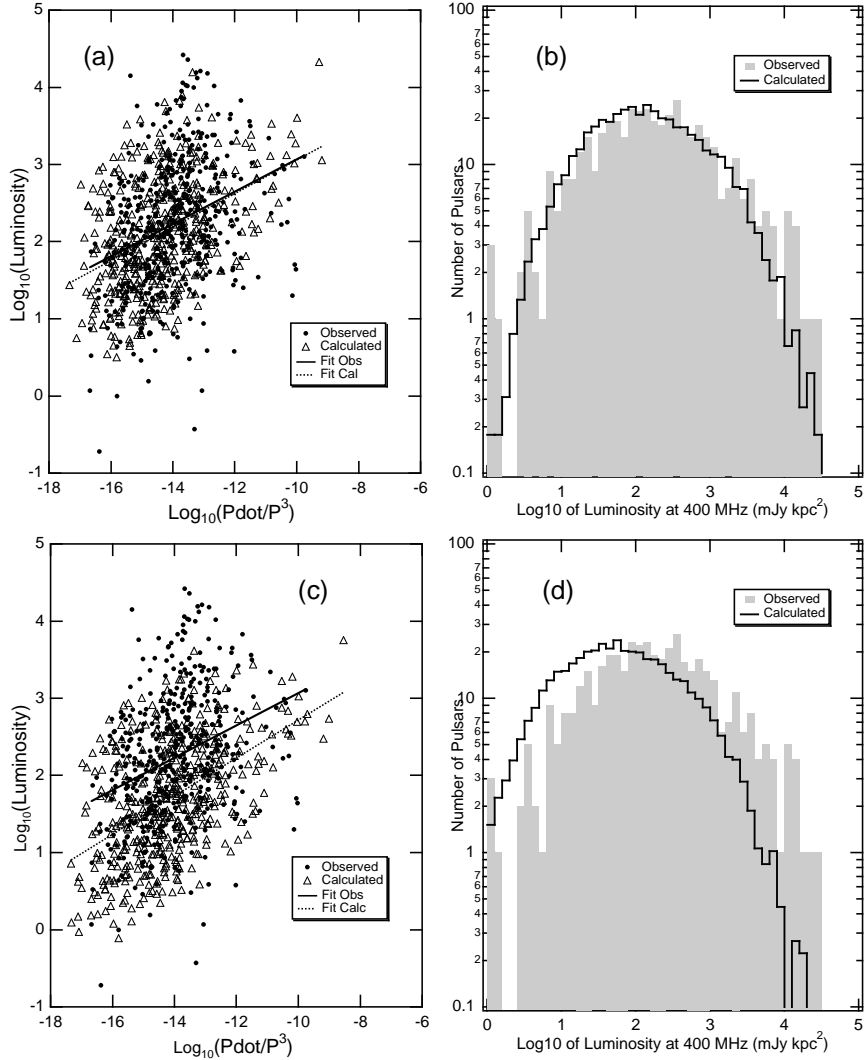} 
\caption{The logarithm of the radio luminosity as a function of the $\log(\dot P/P^3)$
 for the cases of an intercept of 7.2 (a) and 6.64 (c).  Pulsars are simulated assuming
 no magnetic field decay and the presence of a death valley between the death lines.
Observed and simulated pulsars represented by solid dots and open triangles,
respectively.  Linear fits of the observed and simulated distributions are
represented by solid and dotted lines, respectively.  The distributions of the
logarithm of the radio luminosity at 400 MHz for observed (shaded histogram) and
predicted (plain histograms) pulsars for the cases of an intercept of 7.2 (b) 
and 6.64 (d).  {\bf We find that similar results are obtained assuming field
decay.}}
\end{figure}
\begin{figure}
\epsscale{0.75} 
\plotone{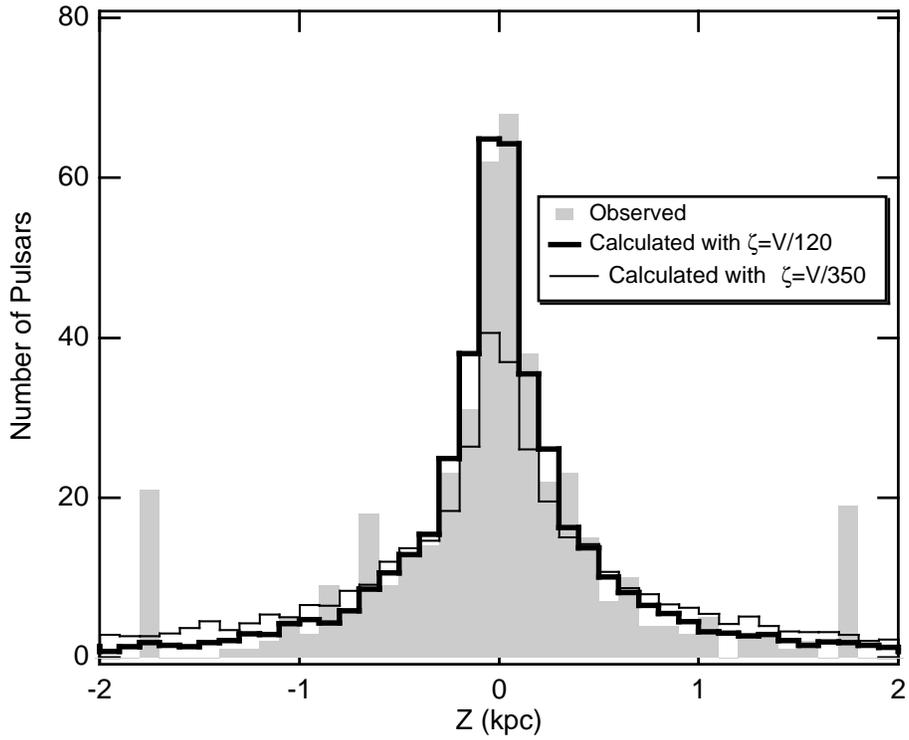} 
\caption{The $z$ distribution
from the Galactic disk for observed (shaded histograms) and simulated
pulsars for the cases of $\zeta=v/120$ (thick line) and of $\zeta=v/350$
(thin line) assuming no field decay and a death valley.  {\bf For the case of field
decay, the simulated distribution is slighty narrower with a width of $0.14\pm 0.01\ kpc$.}}
\end{figure}
\begin{figure}
\epsscale{0.75} 
\plotone{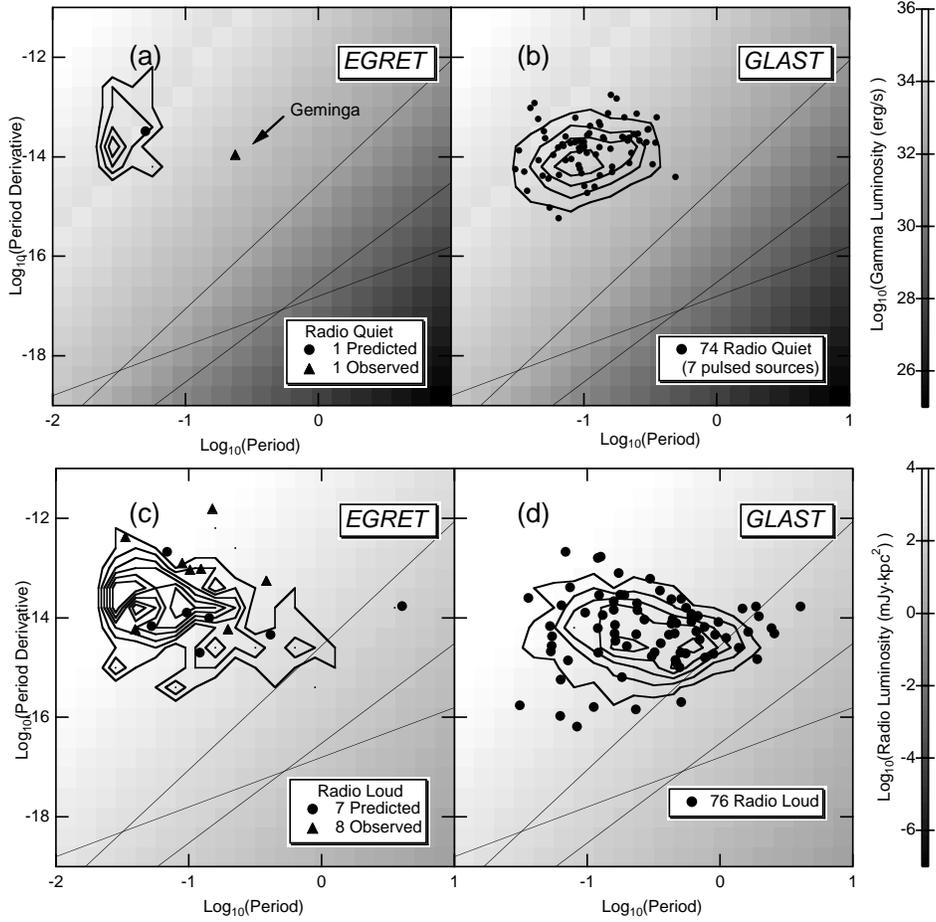} 
\caption{Distribution of radio-quiet (upper panels a and b) and radio-loud 
(lower panels c and d), 
$\gamma$-ray pulsars as a function of period derivative and period for
EGRET and GLAST.  Observed and simulated pulsars are represented by solid
triangles and solid circles, respectively.  The simulated distributions are for the 
case of no magnetic field decay and the presence of a pulsar death valley.
Contours represent the distributions
for a simulation of 10,000 radio pulsars in order to obtain smoother
distributions.  The background gray scale plots represent the average $\gamma$-ray
(upper panels a and b) and radio (lower panels c and d) luminosities.}
\end{figure}
\begin{figure}
\epsscale{0.75} 
\plotone{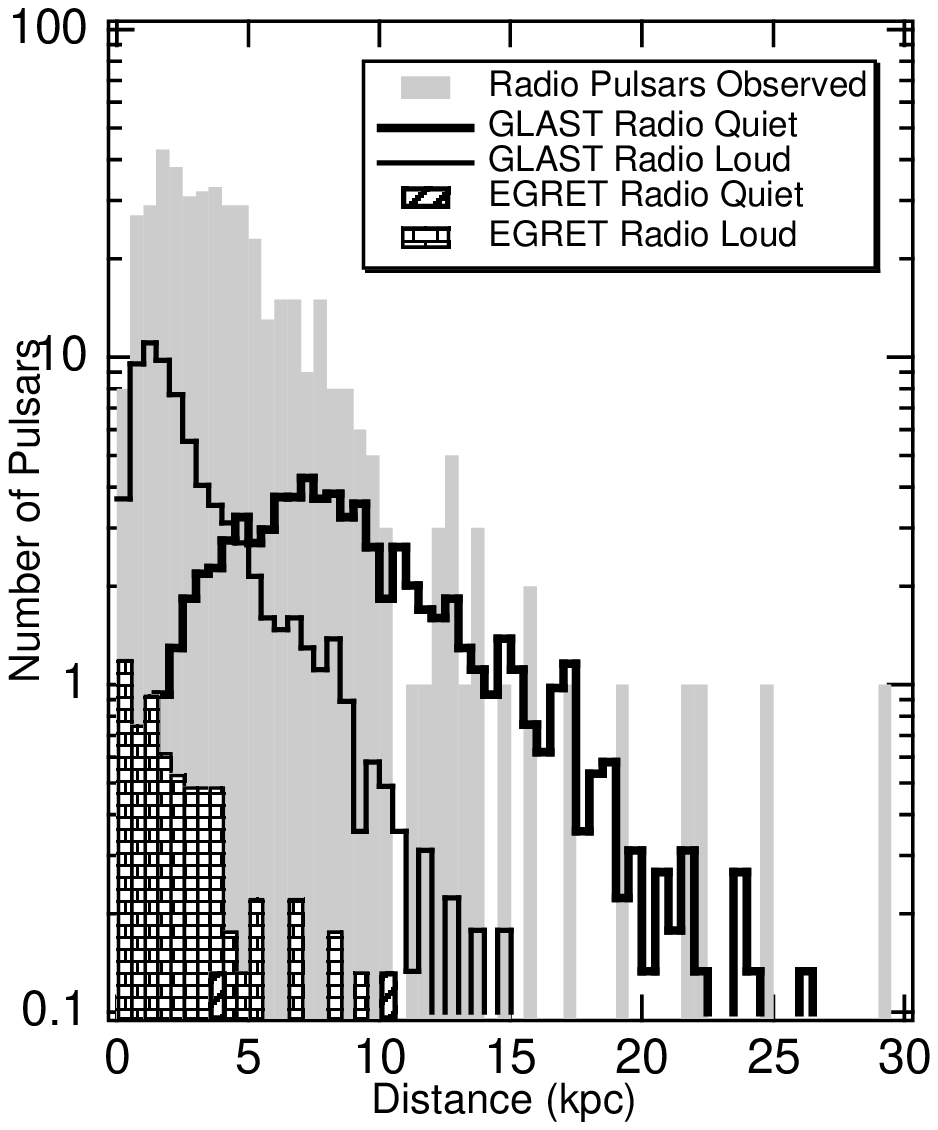} 
\caption{Distributions of the distance from Earth of observed radio pulsars
(shaded histograms), radio-loud (thick histograms) and radio-quiet (thin
histograms), gamma pulsars for GLAST), radio-loud (brick patterned histograms) and
radio-quiet (diagonally patterned histograms), gamma pulsars for EGRET.
The simulated distributions are for the 
case of no magnetic field decay and the presence of a pulsar death valley.}
\end{figure}


\begin{references} 

\reference{}
Arzoumanian, Z., Chernoff, D.F., \& Cordes, J.M., 2001, ApJ submitted.
\reference{}
Bailes, M. 1989, ApJ, 342, 917.
\reference{}
Baring, M. G. \& Harding, A. K. 2000, AAS HEAD Meeting, Honolulu, HI, 
AAS Bull., 32, 1243.
\reference{}
Baring, M. G. \& Harding, A. K. 2001, ApJ, 547, 929.
\reference{}
Bhattacharya, D., Wijers, R.A.M.J., Hartman, J.W., \& Verbunt, F. 1992, A\&AS, 254, 198.
\reference{}
Binney, J., \& Tremaine, S. 1987, {\it Galactic Dynamics}, (Princeton University 
Press, New Jersey), 89.
\reference{}
Boyd, P.T. et al. 1995, ApJ, 448, 365.
\reference{}
Chen, K., \& Ruderman, M.A. 1993, ApJ, 402, 264.
\reference{}
Cheng, K.~S., Ho, C., \& Ruderman, M.~A. 1986, ApJ, 300, 500.
\reference{}
Cheng, K. S., \& Zhang, L. 1998, ApJ, 498, 327.
\reference{}
Clifton, T.R. \& Lyne, A.G. 1986, Nature, 320, 43.
\reference{}
Clifton, T.R, Lyne, A.G., Jones, A.W., McKenna, J., \& Ashworth, M. 1992, MNRAS, 254, 177.
\reference{}
Cordes, J.M., \& Chernoff, D.F. 1998, ApJ, 505, 315.
\reference{}
Daugherty, J.~K., \& Harding A.~K. 1996, ApJ, 458, 278.
\reference{}
Dewey, R.J., \& Cordes, J.M. 1997, ApJ, 321, 780.
\reference{}
Dewey, R.J., Taylor, J.H., Weisberg, J.M., \& Stokes, G.H. 1985, ApJ, 294, L25.
\reference{}
Gehrels, N. et al. 2000, Nature, 404, 363.
\reference{}
Gil, J.A., \& Han, J.L. 1996, ApJ, 458, 265.
\reference{}
Goldreich, P. \& Reisenegger, A. 1992, ApJ, 395, 250.
\reference{}
Grenier, I.A. \& Perrot, C. 1999, Proc. XXVI Int. Cosmic Ray Conf. Salt Lake City, 3, 476.
\reference{}
Gunn, J.E., \& Ostriker, J.P. 1970, ApJ, 160, 979.
\reference{}
Hansen, B.M.S. \& Phinney, E.S. 1997, MNRAS, 291, 569.
\reference{}
Harding, A. K. \& Daugherty, J. K. 1999, A\&AS, 120, 107.
\reference{}
Harding, A. K., \& Muslimov, A. G. 1998, ApJ, 508, 328.
\reference{}
Harding, A. K., \& Muslimov, A. G. 2001, ApJ, 556, 987.
\reference{}
Harding, A.K., \& Zhang, B. 2001, ApJ, 548, L37.
\reference{}
Hartman, J.W. et al. 1997, A\&A, 322, 477.
\reference{}
Hartman, R. et al. 1999, ApJ Supp., 123, 79.
\reference{}
Haslam, C.G.T., Salter, C.J., Stoffel, H., \& Wilson, W.E. 1982, A\&AS, 47,1.
\reference{}
Hirotani, K. \& Shibata, S. 1999, MNRAS, 308, 67.
\reference{}
Johnston, S., Lyne, A.G., Manchester, R.N., Kniffen, D.A., D'Amico, N., Lim, J., \& Ashworth,
M. 1992, MNRAS, 255, 401.
\reference{}
Kaspi, V.M. et al. 1994, ApJ, 422, L83.
\reference{}
Kijak, J., \& Gil, J. 1998, MNRAS, 299, 855.
\reference{}
Kuzmin, A.D. \& Losovsky, B. Ya. 1997, IAU Circular, 6559.
\reference{}
Lorimer, D.R., Bailes, M., \& Harrison, P.A. 1997, MNRAS, 289, 592.
\reference{}
Lyne, A.G. \& Graham-Smith, F., 1998, {\it Pulsar Astronomy}, (Cambridge University 
Press New York), 9.
\reference{}
Lyne, A.G. et al. 1998, MNRAS, 295, 743.
\reference{}
Lyne, A.G., \& Lorimer, D.R. 1994, Nature, 369, 127.
\reference{}
Lyne, A.G., Pritchard, R.S., \& Smith, F.G. 1988, MNRAS, 233, 667.
\reference{}
Malofeev, V.M. \& Malov, O.I. 1997, Nature, 389, 697.
\reference{}
Manchester, R.N. et al. 2001, MNRAS, in press.
\reference{}
Manchester, R.N. et al. 1978, MNRAS, 185, 409.
\reference{}
Manchester, R.N. et al. 1996, MNRAS, 279, 1235.
\reference{}
McLaughlin, M. A. \& Cordes, J. M. 2000, ApJ, 538, 818.
\reference{}
Mollerach, S. \& Roulet, E., 1997, \apj, 479, 147.
\reference{}
Muslimov, A.G., \& Harding, A.K. 1997, ApJ, 485, 735.
\reference{}
Muslimov, A.G., \& Tsygan, A.I. 1992, MNRAS, 255, 61.
\reference{}
Narayan, R., \& Ostriker, J.P. 1990, ApJ, 352, 222.
\reference{}
Nice, D.J., Taylor, J.H., \& Fruchter, A.S. 1993, ApJ, 402, L49.
\reference{}
Paczy\'{n}ski, B. 1990, ApJ, 348, 485.
\reference{}
Press, W.H. et al. 1992, {\it Numerical Recipes in C The Art of Scientific Computing}, (Cambridge
University Press New York), 136.
\reference{}
Pr\'{o}szy\'{n}ski, M., \& Przybyci\'{e}n, D. 1984, {\it in Birth and Evolution of Neutron Stars: Issues Raised
by Millisecond Pulsars}, ed. S.P. Reynolds and D.R. Stinebring (Green Bank: NRAO),
p.151.
\reference{}
Rankin, J.M. 1993, ApJ, 405.
\reference{}
Romani, R.~W. 1996, \apj, 470, 469.
\reference{}
Romani, R. W., \& Yadigaroglu, I.-A. 1995, ApJ, 438, 314.
\reference{}
Shapiro, S. L., \& Teukolsky, S. A., 1983, {\it Black Holes, White Dwarfs, and Neutron Stars The
Physics of Compact Objects} (John Wiley \& Sons New York), 278.
\reference{}
Stokes, G.H., Segelstein, D.J., Taylor, J.H. \& Dewey, R.J.1986, ApJ, 311, 694.
\reference{}
Sturner, S. J., Dermer, C.~D. \& Michel, F.~C. 1995, \apj, 445, 736.
\reference{}
Sturner, S. J., \& Dermer, C.~D. 1996a, \aaps, 120, 99.
\reference{}
Sturner, S. J., \& Dermer, C.~D. 1996b, \apj, 461, 872.
\reference{}
Tauris, T. M. \& Konar, S. 2001, astro-ph/0101531.
\reference{}
Taylor, J.H., \& Cordes, J.M. 1993, ApJ, 411, 674.
\reference{}
Taylor, J.H., Manchester, R.N., \& Lyne, A.G. 1993, ApJS, 88, 529.
\reference{}
Thompson, D. J., Harding, A.K., Hermsen, W., \& Ulmer, M., 1997, 
in Proc. of 4th Compton Symposium, ed. C.D. Dermer, M.S. Strickman \& 
J. Kurfess, p. 39.
\reference{}
Usov, V. V., \& Melrose, D. B. 1995, Aust. J. Phys., 48, 571.
\reference{}
van den Heuvel, E. P. J. \& van Paradijs, J. 1997, ApJ, 483, 399.
\reference{}
Vivekanand, M., \& Narayan, R. 1981, J. Ap. Astr., 2, 315.
\reference{}
Yadigaroglu, I. A. \& Romani, R. W. 1995, ApJ, 449, 211.
\reference{}
Zhang, B., \& Harding, A. K. 2000, ApJ, 532, 1150.
\reference{}
Zhang, B., Harding, A. K., \& Muslimov, A. G. 2000, ApJ, 531, L135.
\reference{}
Zhang, L., Zhang, Y. J. \& Cheng, K. S. 2000, A \& A, 357, 957.

\end{references}
\end{document}